\documentclass[10pt,twocolumn,journal]{IEEEtranw}
\usepackage[left=1.2cm,top=1.5cm,right=1.2cm,bottom=1.7cm]{geometry}
\usepackage{url,amsmath,graphicx}
\usepackage{amsfonts}
\usepackage{multirow}
\usepackage[T1]{fontenc}
\usepackage{bm}
\usepackage{relsize}
\usepackage{pifont}
\usepackage{color}
\usepackage{textcomp}
\usepackage{wasysym}
\usepackage{stmaryrd}
\usepackage{amssymb}

\DeclareMathOperator{\argmin}{argmin}
\begin{document}
\title{ Angular Parameters Estimation of Multiple Incoherently Distributed
Sources Generating Noncircular Signals} \author{Sonia Ben Hassen, Faouzi
Bellili, Abdelaziz Samet, and Sofi\`ene Affes \\\small INRS-EMT, 800, de la Gaucheti\`ere Ouest, Bureau 6900, Montreal, Qc, H5A 1K6, Canada
\\\small Emails:sonia.benhassen@ieee.org, bellili@emt.inrs.ca, abdelaziz.samet@ept.rnu.tn, affes@emt.inrs.ca
\vspace{0.3cm}
\thanks{Work supported by the Discovery Grants Program of NSERC and a Discovery Accelerator Supplement Award from NSERC.}
}

\maketitle
\begin{abstract}
We introduce a new method for the estimation of the angular parameters [i.e., central directions of arrival (DOAs) and angular spreads] of multiple non-circular and incoherently-distributed (ID) sources and thoroughly analyze its performance. By decoupling the estimation of the central DOAs from that of the angular spreads, we reduce significantly the complexity of the proposed technique. The latter outperforms most well-known state-of-the-art techniques in terms of estimation accuracy and robustness.
\end{abstract}
\section{Introduction}
Direction of arrivals estimation for multiple plane waves impinging on an arbitrary array of sensors has received a significant amount of
attention over the last several decades [\ref{LEEE01}].
It has typically found many applications in different areas such as modern wireless communication
systems [\ref{LEEE1}], audio/speech processing systems [\ref{LEEE2}], radar and sonar [\ref{LEEE3}], just to name a few.
In most applications, however, DOA estimation
methods are based on the point-source model which postulates that the signals are generated from far-field point sources and travel along a single path to the receiving antenna array.
Using this simplified model, many DOA estimators have been developed for both temporally uncorrelated [\ref{LEEE4}-\ref{LEEE6}] and correlated [\ref{LEEE7},\ref{LEEE8}] signals. 
However, in real-world surroundings, especially in typical urban environments, multipath propagation made by a cluster of reflections close to each mobile causes angular spreading [\ref{LEEE9}]. In other words, the signal radiated by each source hits the antenna array via different paths with different angles. In this more
realistic model, the source is viewed by the array as spatially distributed, i.e., with a central  DOA and an angular spread. The latter influences the quality of the communication link and represents an important characteristic
for spatial diversity schemes [\ref{LEEE10}, \ref{LEEE11}]. DOA estimation
becomes more challenging in presence of local scattering [\ref{LEEE12}, \ref{LEEE13}] because the latter affects the signal spatial distribution. In this context, some studies have shown that classical point-source estimation methods suffer from severe performance degradation when applied to the \textit{distributed-source} scenario [\ref{LEEE14}, \ref{LEEE15}]. This observation has prompted an increasing interest, over the few recent years, in developing  DOA estimation algorithms that can handle
both point and scattered sources in order to improve direction finding capabilities in real-world propagation environments.
\\ Depending on the nature of scattering, signal components arriving from different directions exhibit varying degrees of correlation. Hence, we distinguish two different types of the propagation channel. The first one is when the received
signal components originated from a source and scattered at different angles are delayed and scaled replicas of the same signal. This feature is known in the literature as ``coherent source distribution'' or ``coherently-distributed (CD) source'' [\ref{LEEE16}]. The second type of the propagation channel corresponds to the fact that the signal components of a source impinging from different scatterers at different angles are uncorrelated. This is termed in the literature as ``incoherent source distribution'' or ``incoherently-distributed (ID) source'' [\ref{LEEE16}, \ref{LEEE17}]. Therefore, for uncorrelated CD sources, each source contributes rank-one component to the spatial covariance matrix and, as such, the rank of the noise-free covariance
matrix is equal to the number of sources [\ref{LEEE16}]. Consequently, many
classical DOA estimation methods based on the simplistic point-source model can be easily extended to CD sources. Particulary, authors proposed in [\ref{LEEE171}] an efficient DSPE algorithm for estimating the angular parameters of CD sources. This method enables a decoupled estimation of the DOAs from that of the angular spreads of sources with small angular spread.
However, for ID sources, the whole observation space is occupied by signal components, and the noise subspace is generally degenerate [\ref{LEEE16}]. Therefore, the rank of the noise-free
covariance matrix is different from the number of sources; it even increases with the angular spread. This makes the trivial generalization of traditional point-source subspace-based methods to the ID case not feasible.
To sidestep this problem, tremendous efforts have been directed to developing  new angular parameters estimators  that are specifically tailored to ID sources. In particular, techniques that are able to handle a single ID source were developed in [\ref{LEEE18}-\ref{LEEE24}].\\
Many estimators were also developed to estimate the angular parameters of  multiple ID sources. In fact, a class of subspace methods were proposed in [\ref{LEEE9}, \ref{LEEE16}, \ref{LEEE17}] wherein the effective dimension of the signal subspace is defined as the number of the first eigenvalues (of the noise-free
covariance matrix) that reflect most of the signal energy.
 More computationally attractive approaches that are based on the beamforming techniques were later introduced in [\ref{LEEE25}, \ref{LEEE26}]. Despite their good performance, all these estimators assume the angular distributions to be perfectly \textit{known} and \textit{identical} to all the sources. Methods which are able to handle the multi-source
case with \textit{known} but \textit{different} angular distributions were also proposed in [\ref{LEEE27}, \ref{LEEE28}]. Recently, a robust version of the generalized Capon principle [\ref{LEEE26}] (RGC) has been developed in [\ref{LEEE29}] which, in contrast to all existing approaches, does not need the \textit{a priori} knowledge of the  angular distributions. Moreover, the latter does not need  to be the same for all the sources. This robust approach is, however, statistically less efficient than the aforementioned subspace-based (high-resolution) methods [\ref{LEEE9}, \ref{LEEE16}, \ref{LEEE17}], especially in the presence of closely-spaced ID sources.
\par More recently, A. Zoubir \textit{et al.} proposed an efficient subspace-based (ESB) algorithm [\ref{LEEE30}] to estimate the angular parameters of multiple ID \textit{circular} sources. ESB enjoys a good trade-off between estimation performance and computational complexity. In order to alleviate the computational burden stemming from eigendecomposing the covariance matrix, ESB  exploits the properties of its inverse and estimates the angular parameters using a 2-D search.
Both the statistical efficiency and high-resolution capabilities of the subspace-based techniques are maintained and, most interestingly, ESB is not limited to a particular antenna array geometry or to a specific type of scatterers' angular distribution.
Yet, it still requires the angular distribution to be perfectly \textit{known} and \textit{identical} for all the sources on the top of being derived specifically for \textit{circular} sources.
Recently, a new method for tracking the central DOAs assuming multiple ID mobile sources has been also proposed in [\ref{LEEE301}]. It is based on a simple covariance fitting optimization
technique [\ref{LEEE28}] to estimate the
central DOAs and the Kalman filter to model the dynamic property of directional changes for the moving sources. Despite its efficiency, this method requires the sources' angular distributions to be perfectly \textit{known} and is derived for \textit{circular} sources only.
\par Noncircular signals, however, such as binary-phase-shift-keying (BPSK) and offset quadrature-phase shift-keying (OQPSK)-modulated signals, are also frequently encountered in digital communications. Therefore, there has been a recent surge of interest in deriving new algorithms that are able to properly handle \textit{noncircular} signals as well [\ref{LEEE3011}-\ref{LEEE347}]. These estimators extract additional information about the angular parameters from the \textit{unconjugated} spatial covariance matrix that is non-zero for \textit{noncircular} sources, in contrast to \textit{circular} ones. From this perspective, we have been also able to propose a robust technique which is able to handle both temporally and spatially correlated sources in presence of noncircular signals [\ref{LEEE348}]. By accounting for both signals' noncircularity and temporal correlation, the proposed estimator was indeed shown to offer huge performance enhancements with respect to the main state-of-the-art techniques. Yet, all the aforementioned estimators [\ref{LEEE31}-\ref{LEEE348}] are applicable for the point-source model  only. And, to the best of the authors' knowledge, no contribution has dealt so far with the problem of angular parameters estimation (i.e., central DOAs and angular spreads) of multiple \textit{noncircular} ID sources.
\par Motivated by these fact, we tackle in this paper for the very first time the problem of estimating the angular parameters of ID \textit{noncircular} sources.
We propose a new method that allows decoupling the estimation of each central DOA from its associated angular spread in the presence of \textit{noncircular} sources. This method will be derived by going through three different stages resulting in two versions of the proposed estimator. The first one is a new 2-D search algorithm that extends ESB from \textit{circular} to \textit{noncircular} sources. And the second is a robust version that estimates the angular parameters by means of two successive one-dimensional (1-D) parameter searches. Towards this goal, we will use unstructured models for the \textit{conjugated} and \textit{unconjugated} noise-free covariance matrices that depend on the unknown angular spreads only. Most interestingly, such unstructured models are totally oblivious to the angular distributions of the sources and, therefore, their \textit{a priori} knowledge is not  required by the proposed method; a quite precious degree of freedom in practice. Even more, unlike all the existing methods, the proposed technique does not need to assume the same angular distribution across all the sources.\\
In order to properly assess the performance of the new estimator, we also conduct a complete theoretical study of its statistical properties (i.e., its bias and variance). Furthermore, we derive an explicit expression for the CRLB of the underlying estimation problem. This fundamental lower bound, which reflects the best achievable performance ever [\ref{LEEE349}], will be used as an overall benchmark against which we gauge the accuracy of the new estimator. Computer simulations will show that the proposed estimator outperforms ESB and RGC especially at low SNR values and/or low DOA separations.
The new CRLBs will also reveal that the noncircularity of the signals becomes more informative about the angular
parameters when the sources have different angular
distributions and when the angular spreads increase.
\par The rest of this paper is organized as follows. In Section \ref{section_2}, we introduce the system model and some of the basic assumptions that will be adopted throughout the
article. In section \ref{section_3}, we derive the new algorithm and in section \ref{section_31} we show how the
estimation of the central DOAs can be decoupled from that of the
angular spreads. In section \ref{section_4}, we derive the statistical bias and variance of the new estimator. In section \ref{section_5},  we derive an explicit expression for the CRB of the underlying estimation problem. Computer simulations are presented in Section  \ref{section_6} and concluding remarks are drawn out in Section \ref{section_7}.\\
We list beforehand  some of the common notations adopted throughout this paper. Matrices and vectors are represented by bold upper- and lower-case characters, respectively. Vectors are by default in
column orientation. Moreover, we consider the following standard notations:
\small
\begin{eqnarray}
\delta(.)&:& \textrm{Dirac delta function}\nonumber;\\
(.)^*&:&\textrm{Complex conjugate};\nonumber\\
\angle(.)&:&\textrm{Phase angle (or argument) in radians};\nonumber\\
|.|&:&\textrm{Complex modulus};\nonumber\\
(.)^T&:&\textrm{Transpose};\nonumber\\
(.)^H&:&\textrm{Conjugate transpose};\nonumber\\
\simeq&:&\textrm{Approximately equal};\nonumber\\
\textrm{argmin}_K\{.\}&:& \textrm{Position of the $K$ minima of any given}\nonumber\\
&& \textrm{ objective function};\nonumber\\
\textrm{tr}\{\mathbf{A}\}&:&\textrm{Trace of a given matrix}~ \mathbf{A};\nonumber\\
\textrm{diag}\{\bm{v}\}&:&\textrm{Diagonal matrix whose main diagonal's }\nonumber\\
&&\textrm{elements are those of vector}~ \bm{v};\nonumber\\
\left\|.\right\|_{\textrm{Fro}}&:&\textrm{Frobenius norm};\nonumber\\
\Re\{.\}&:&\textrm{Real part operator};\nonumber\\
\mathbb{E}\{.\}&:&\textrm{Statistical expectation};\nonumber\\
\frac{\partial^{n}(.)}{\partial(.)^n}&:&\textrm{$n^{th}$-order partial derivative};\nonumber\\
\textrm{eig}\{\mathbf{A}\}&:&\textrm{Eigenvalues of a matrix}~\mathbf{A};\nonumber
\end{eqnarray}
\begin{eqnarray}
\varodot&:&\textrm{Hadamard-Schur product};\nonumber\\
\mathbf{I}_p&:&\textrm{$(p\times p)$ identity matrix};\nonumber\\
\bm{0}_{p\times q}&:&\textrm{$(p\times q)$ zero matrix};\nonumber\\
\textrm{Toeplitz}\{\bm{v}\}&:& \textrm{Symmetric Toeplitz matrix constructed  }\nonumber\\
&&\textrm{from a given vector}~ \bm{v};\nonumber\\
\textrm{Hankel}\{ \bm{v}_1, \bm{v}_2\}&:&\textrm{Hankel matrix constructed from  the}\nonumber\\
&&\textrm{vectors}~ \bm{v}_1~\textrm{and} ~  \bm{v}_2;\nonumber\\
a_l(\theta) &:& \textrm{Response of
the $l$th sensor to a unit-energy}\nonumber\\
 &&\textrm{source radiating from direction}~ \theta;\nonumber\\
{f_l(\theta)}_{l=0,2,\ldots,(L-1)}&:& \textrm{Real-valued transformations of the scalar}\nonumber\\
&&\textrm{DOA parameter}~ \theta;\nonumber\\
\bar{\Theta}_k &:& \textrm{Central DOA of each $k^{th}$ source};\nonumber\\
\bar{\sigma}_k &:& \textrm{Angular spread of each $k^{th}$ source};\nonumber\\
\rho_k(\theta,\bar{\bm{\psi}}_k) &:& \textrm{\textit{Normalized} angular
power density}\nonumber\\
&&\textrm{of the $k$th source};\nonumber\\
p_{kk}(\theta, \theta';\bar{\bm{\psi}}_k) &:& \textrm{\textit{Conjugated} angular auto-correlation kernel}\nonumber\\
&&\textrm{of the $k^{th}$ source};\nonumber\\
p'_{kk}(\theta, \theta';\bar{\bm{\psi}}_k) &:& \textrm{\textit{Unconjugated} angular auto-correlation kernel}\nonumber\\
&&\textrm{of the $k^{th}$ source};\nonumber\\
p_{kk'}(\theta, \theta';\bar{\bm{\psi}}_k,\bar{\bm{\psi}}_{k'})&:& \textrm{\textit{Conjugated} angular cross-correlation kernel}\nonumber\\
&&\textrm{between sources $k$ and $k'$};\nonumber\\
p'_{kk'}(\theta, \theta';\bar{\bm{\psi}}_k,\bar{\bm{\psi}}_{k'}) &:& \textrm{ \textit{Unconjugated} angular cross-correlation}\nonumber\\
&&\textrm{kernel between sources $k$ and $k'$}.\nonumber
\end{eqnarray}
\normalsize
\section{System Model}\label{section_2}
Consider an array consisting of $L$ identical sensors (i.e., with the same gain, phase, and sensitivity pattern) that is immersed in the far-filed of $K$ scattered ID sources with the same central frequency $\omega_0$. Assume that the root mean square (rms) delay spread is small compared to the inverse bandwidth of the transmitted signals so that the narrowband assumption remains valid in the presence of scattering [\ref{LEEE4190}-\ref{LEEE41902}]. Under these mild conditions, the signal received by the $l$th sensor, $l=1,2,\ldots,L$, can be modeled as follows [\ref{LEEE26}-\ref{LEEE30}]:
\begin{eqnarray}\label{equ1}
x_l(n)&=&\displaystyle\sum_{k=1}^{K}\int a_l(\theta)s_k(\theta, \bar{\bm{\psi}}_k,n)d\theta~+~w_{l}(n),
\end{eqnarray}
in which $n$ stands for the $n$th snapshot. Moreover, $w_{l}(n)$ is an additive zero-mean circularly symmetric Gaussian-distributed noise. The noise components are assumed to be temporally and spatially white, i.e., uncorrelated between snapshots and receiving antenna branches, respectively. Furthermore, $s_k(\theta,\bar{\bm{\psi}}_k,n)$ is the data-modulated angular distribution (with respect to $\theta$) of the signal received from the $k$th source; parameterized here by the vector $\bar{\bm{\psi}}_k=[\bar{\Theta}_k,\bar{\sigma}_k]^T$.\\
For any planar configuration of the
receiving antenna array, $a_l(\theta)$ can be written as:
\begin{eqnarray}
a_l(\theta)&=&e^{j2\pi f_{l-1}(\theta)}.
\end{eqnarray}
For mathematical convenience, we gather all the unknown \textit{central} DOAs and angular spreads in the following  parameter vectors:
\begin{eqnarray}
\label{centrdoa}\bar{\bm{\Theta}}&\triangleq&\left[\bar{\Theta}_1,\bar{\Theta}_2,\ldots,\bar{\Theta}_K\right]^T,\\
\label{angspre}\bar{\bm{\sigma}}&\triangleq&\left[\bar{\sigma_1},\bar{\sigma_2},\ldots,\bar{\sigma}_K\right]^T.
\end{eqnarray}
Our goal in the remainder of this paper is to jointly estimate the angular parameters, $\bar{\bm{\Theta}}$ and $\bar{\bm{\sigma}}$, of the $K$ noncircular sources given the set of received signals, $x_l(n)$, $l=1,2,\ldots,L$. To that end, we stack the received data over the $L$ sensors at each snapshot $n$ in a single vector:
\begin{eqnarray}
\mathbf{x}(n)&\triangleq&[x_1(n),\ldots,x_L(n)]^T.
\end{eqnarray}
From (\ref{equ1}), $\mathbf{x}(n)$ is explicitly given by:
\begin{eqnarray}\label{equ2}
\mathbf{x}(n)&=&\displaystyle\sum_{k=1}^{K}\int\mathbf{a}(\theta)s_k(\theta,\bar{\bm{\psi}}_k,n)d\theta~+~\mathbf{w}(n),
\end{eqnarray}
where
\begin{eqnarray}
\mathbf{w}(n)&\triangleq&\left[w_{1}(n),\ldots,w_{L}(n)\right]^T,\nonumber\\
\mathbf{a}(\theta)&\triangleq&\left[a_1(\theta),\ldots,a_L(\theta)\right]^T,\nonumber
\end{eqnarray}
are the array noise and  response vectors, respectively. 
For ID sources, the components impinging from different scatterers are uncorrelated thereby yielding:
\begin{eqnarray}
\label{eqpkkps}p_{kk}(\theta, \theta';\bar{\bm{\psi}}_k)&\triangleq&\mathbb{E}\Big\{s_k\big(\theta,\bar{\bm{\psi}}_k,n\big)~\!s_k\big(\theta',\bar{\bm{\psi}}_k,n\big)^{*}\Big\},\\
\label{psi2}&=&\sigma_{s_k}^2\rho_k(\theta,\bar{\bm{\psi}}_k)\delta(\theta-\theta'),
\end{eqnarray}
where  $\sigma_{s_k}^2$ is the average power of the $k$th source.
Since the sources are also assumed to radiate \textit{noncircular} signals, we adopt the definition of noncircularity in [\ref{LEEE3491}, \ref{LEEE3492}]. Moreover, we exploit the property of signals' correlation in the real sense [\ref{LEEE3492}, property $3.1$] to prove from (\ref{psi2}) that $p'_{kk}(\theta, \theta';\bar{\bm{\psi}}_k)$ can be written as:
\begin{eqnarray}
\label{unconjugated_kernels}p'_{kk}(\theta, \theta';\bar{\bm{\psi}}_k)&\!\!\!\triangleq\!\!\!&\mathbb{E}\Big\{s_k\big(\theta,\bar{\bm{\psi}}_k,n\big)~\!s_k\big(\theta',\bar{\bm{\psi}}_k,n\big)\Big\},\\
\label{second_corr}&\!\!\!=\!\!\!&\sigma_{s_k}^2\bar{\gamma}_k e^{j\bar{\varphi}_k}\rho_k(\theta,\bar{\bm{\psi}}_k)\delta(\theta-\theta').
\end{eqnarray}
Here, $0\leq\bar{\gamma}_k\leq 1$ and $\bar{\varphi}_k$ are the noncircularity rate and phase of the $k$th source, respectively.  As emphasized in Section II, all existing works on angular parameters estimation of ID sources assume the sources to be \textit{circular}. As such, none of them makes use of the \textit{unconjugated} kernels in (\ref{unconjugated_kernels})  since they are identically zero in this case.
In this paper, however, we consider the case of \textit{noncircular} sources with maximum noncircularity rate (i.e., $\bar{\gamma}_k=1$), known in the open literature as strictly second-order \textit{noncircular} or \textit{rectilinear} signals. Examples of such signals include  unfiltered BPSK-, OQPSK-, PAM-, ASK-, AM- and MSK-modulated signals [\ref{LEEE32}].
Their  \textit{unconjugated} angular auto-correlation kernels are obtained from (\ref{second_corr}) as:
\begin{eqnarray}\label{kernelprim}
p'_{kk}(\theta, \theta';\bar{\bm{\psi}}_k)&=&\sigma_{s_k}^2 e^{j\bar{\varphi}_k}\rho_k(\theta,\bar{\bm{\psi}}_k)\delta(\theta-\theta').
\end{eqnarray}
Now, since the sources' signals are uncorrelated from the noise components, the \textit{conjugated} and \textit{unconjugated} covariance matrices of $\mathbf{x}(n)$ defined, respectively, as $\mathbf{R}_{\mathbf{x}\mathbf{x}}=\mathbb{E}\{\mathbf{x}(n)\mathbf{x}(n)^H\}$ and $\mathbf{R}'_{\mathbf{x}\mathbf{x}}=\mathbb{E}\{\mathbf{x}(n)\mathbf{x}(n)^T\}$ are explicitly given by:
\begin{eqnarray}
\label{covmat}\!\!\mathbf{R}_{\mathbf{x}\mathbf{x}}&\!\!\!\!\!=\!\!\!\!\!&
\displaystyle\sum_{k=1}^{K}\displaystyle\sum_{k'=1}^{K}\!\int\!\!\!\!\int p_{kk'}(\theta, \theta';\bar{\bm{\psi}}_k,\bar{\bm{\psi}}_{k'})\mathbf{a}(\theta)\mathbf{a}^H(\theta')d\theta d\theta'~\!\!\!+~\!\!\!\sigma_{w}^2\mathbf{I}_L,\nonumber\\
&\!\!\!\!\!\!\!\!\!\!&\\
\label{covmatprim}\!\!\mathbf{R}'_{\mathbf{x}\mathbf{x}}&\!\!\!\!\!\!=\!\!\!\!\!\!&\displaystyle\sum_{k=1}^{K}\displaystyle\sum_{k'=1}^{K}\!\int\int p'_{kk'}(\theta, \theta';\bar{\bm{\psi}}_k,\bar{\bm{\psi}}_{k'})\mathbf{a}(\theta)\mathbf{a}^T(\theta')d\theta d\theta'\!,
\end{eqnarray}
where $\sigma_{w}^2$ is the unknown noise variance. Note here that the \textit{unconjugated} covariance matrix of the \textit{circular} noise vector is identically zero and, therefore, it vanishes in (\ref{covmatprim}) contrarily to (\ref{covmat}).
By further assuming the ID sources to be mutually uncorrelated, it follows that:
\begin{eqnarray}
\label{psi}p_{kk'}(\theta, \theta';\bar{\bm{\psi}}_k,\bar{\bm{\psi}}_{k'})&=&p_{kk}(\theta, \theta';\bar{\bm{\psi}}_k,\bar{\bm{\psi}}_k)\delta_{kk'},\\
\label{psiprim}p'_{kk'}(\theta, \theta';\bar{\bm{\psi}}_k,\bar{\bm{\psi}}_{k'})&=&p'_{kk}(\theta, \theta';\bar{\bm{\psi}}_k,\bar{\bm{\psi}}_k)\delta_{kk'},
\end{eqnarray}
where $\delta_{kk'}$ is the Kronecker delta function defined as $\delta_{kk'}=1$ for $k=k'$ and $0$ otherwise. Now, plugging (\ref{psi2}) and (\ref{kernelprim}) in (\ref{psi}) and (\ref{psiprim}), respectively, leads to:
\begin{eqnarray}
\!\!\!\!\label{expkk'}p_{kk'}(\theta,\theta';\bar{\bm{\psi}}_k,\bar{\bm{\psi}}_{k'})&\!\!=\!\!&\sigma_{s_k}^2\rho_k(\theta,\bar{\bm{\psi}}_k)\delta(\theta-\theta')\delta_{kk'},\\
\!\!\!\!\label{expprimkk'}p'_{kk'}(\theta,\theta';\bar{\bm{\psi}}_k,\bar{\bm{\psi}}_{k'})&\!\!=\!\!&\sigma_{s_k}^2 e^{j\bar{\varphi}_k}\rho_k(\theta,\bar{\bm{\psi}}_k)\delta(\theta-\theta')\delta_{kk'}.
\end{eqnarray}
Consequently, (\ref{covmat}) and (\ref{covmatprim}) simplify to:
\begin{eqnarray}
\label{matrx}\!\!\!\!\!\!\!\!\!\!\!\mathbf{R}_{\mathbf{x}\mathbf{x}}&\!\!=\!\!&\displaystyle\sum_{k=1}^{K}\int\sigma_{s_k}^2\rho_k(\theta,\bar{\bm{\psi}}_k)\mathbf{a}(\theta)\mathbf{a}(\theta)^Hd\theta~+~\sigma_{w}^2\mathbf{I}_L,\\
\label{secondmatrx}\!\!\!\!\!\!\!\!\!\!\!\mathbf{R}'_{\mathbf{x}\mathbf{x}}&\!\!=\!\!&\displaystyle\sum_{k=1}^{K}\int \sigma_{s_k}^2e^{j\bar{\varphi}_k}\rho_k(\theta,\bar{\bm{\psi}}_k)\mathbf{a}(\theta)\mathbf{a}(\theta)^Td\theta.
\end{eqnarray}
%
 \section{Angular Parameters Estimation in Presence of Noncircular Signals}\label{section_3}
In order to exploit the additional information contained in the \textit{unconjugated} covariance matrix of noncircular signals, we define the following \textit{extended} received vector:
\begin{eqnarray}\label{extrecvect}
\widetilde{\mathbf{x}}(n)&\triangleq&\left[\!\!
\begin{array}{cc}
\mathbf{x}(n)^T~~{\mathbf{x}}(n)^H
\end{array}\!\!\right]^T.
\end{eqnarray}
whose extended covariance matrix is given by:
\begin{eqnarray}\label{exextcomxRxx}
\mathbf{R}_{\widetilde{\mathbf{x}}\widetilde{\mathbf{x}}}&=&\mathbb{E}\Big\{\widetilde{\mathbf{x}}(n)\widetilde{\mathbf{x}}(n)^H\Big\}~=~\left(
\begin{array}{cc}
\!\!\mathbf{R}_{\mathbf{x}\mathbf{x}}&\mathbf{R}'_{\mathbf{x}\mathbf{x}}\!\!\\
\!\!\mathbf{R}'^{*}_{\mathbf{x}\mathbf{x}}&\mathbf{R}_{\mathbf{x}\mathbf{x}}^*\!\!
\end{array}\right).
\end{eqnarray}
On the one hand, using the explicit expressions of $\mathbf{R}_{\mathbf{x}\mathbf{x}}$ and $\mathbf{R}'_{\mathbf{x}\mathbf{x}}$ established,  respectively, in (\ref{matrx}) and (\ref{secondmatrx})  and resorting to some algebraic manipulations, it can be shown that:
\begin{eqnarray}\label{extmatrx}
\mathbf{R}_{\widetilde{\mathbf{x}}\widetilde{\mathbf{x}}}&\!\!\!=\!\!\!&\displaystyle\sum_{k=1}^{K}\int\sigma_{s_k}^2\rho_k(\theta,\bar{\bm{\psi}}_k)~\!\widetilde{\mathbf{a}}(\theta,\bar{\varphi}_k)~\!\widetilde{\mathbf{a}}(\theta,\bar{\varphi}_k)^Hd\theta~+~\sigma_{w}^2\mathbf{I}_{2L},\nonumber\\&&
\end{eqnarray}
where $\widetilde{\mathbf{a}}(\theta,\bar{\varphi}_k)$ is the extended array response vector defined as:
\begin{eqnarray}\label{extended-response}
\widetilde{\mathbf{a}}(\theta,\bar{\varphi}_k)&\triangleq&\left[\!\!
\begin{array}{cc}
\mathbf{a}(\theta)^T,~~e^{-j\bar{\varphi}_k}{\mathbf{a}}(\theta)^H
\end{array}\!\!\right]^T.
\end{eqnarray}
We also define the extended (normalized) covariance matrix of the noise-free signal pertaining to the $k$th source as:
\begin{eqnarray}\label{extended_noise_free}
\!\!\!\!\!\!\!\!\widetilde{\mathbf{R}}^{(k)}_{ss}(\bar{\bm{\psi}}_k,\bar{\varphi}_k)&\!\triangleq\!\!&\int\rho_k(\theta,\bar{\bm{\psi}}_k)\widetilde{\mathbf{a}}(\theta,\bar{\varphi}_k)\widetilde{\mathbf{a}}(\theta,\bar{\varphi}_k)^Hd\theta.
\end{eqnarray}
Hence, the extended covariance matrix in (\ref{extmatrx}) is simply given by:
\begin{eqnarray}\label{exmatrxnv}
\mathbf{R}_{\widetilde{\mathbf{x}}\widetilde{\mathbf{x}}}&=&\displaystyle\sum_{k=1}^{K}~\sigma_{s_k}^2\widetilde{\mathbf{R}}^{(k)}_{ss}(\bar{\bm{\psi}}_k,\bar{\varphi}_k)~+~\sigma_{w}^2\mathbf{I}_{2L}.
\end{eqnarray}
Next, we consider the following eigendecomposition of the extended covariance matrix in (\ref{exmatrxnv}):
\begin{eqnarray}
\label{eigextR}\mathbf{R}_{\widetilde{\mathbf{x}}\widetilde{\mathbf{x}}}&=&\widetilde{\mathbf{U}}_s\bm{\Sigma}\widetilde{\mathbf{U}}_s^H~+~\sigma_{w}^2\widetilde{\mathbf{U}}_w\widetilde{\mathbf{U}}_w^H,
\end{eqnarray}
where
$\widetilde{\mathbf{U}}_s$ and $\widetilde{\mathbf{U}}_w$ denote the eigenvector matrices associated
to the signal and noise subspaces, respectively. Moreover, $\bm{\Sigma}$ is a diagonal matrix containing the eigenvalues of the overall extended noise-free covariance matrix involved in (\ref{exmatrxnv}), i.e.:
\begin{eqnarray}
\mathbf{R}_{\tilde{s}\tilde{s}}&\triangleq&\sum_{k=1}^{K}\sigma_{s_k}^2\widetilde{\mathbf{R}}^{(k)}_{ss}(\bar{\bm{\psi}}_k,\bar{\varphi}_k).
\end{eqnarray}
Traditional subspace-based methods which are all designed for \textit{circular} ID sources rely on the fact that the columns of each $k$th noise-free covariance matrix, $\mathbf{R}^{(k)}_{ss}(\bar{\bm{\psi}}_k)$, are orthogonal to those of the pseudo-noise subspace, i.e.:
\begin{eqnarray}\label{relsubsorig}
\mathbf{U}_w^H~\!\mathbf{R}^{(k)}_{ss}(\bar{\bm{\psi}}_k)&=&\bm{0}_{(L-r)\times L},
\end{eqnarray}
in which $r$ is the effective dimension of the pseudosignal
subspace [\ref{LEEE17}]. In principle, the same orthogonality property in (\ref{relsubsorig}) holds for \textit{noncircular} ID sources:
\begin{eqnarray}\label{relsubs}
\widetilde{\mathbf{U}}_w^H\widetilde{\mathbf{R}}^{(k)}_{ss}(\bar{\bm{\psi}}_k,\bar{\varphi}_k)&=&\bm{0}_{(2L-r)\times 2L},
\end{eqnarray}
and can be used, as well, to estimate the associated angular parameters. However, similar to all subspace methods, the estimation performance is critically affected if the effective dimension $r$ is not appropriately selected. Besides, the optimal choice of $r$ depends on the value of the angular spread which is itself considered as an unknown parameter in our work. To sidestep this problem, we will rather capitalize on the inverse of the extended covariance matrix as recently done in [\ref{LEEE30}]:
\begin{eqnarray}\label{inverseRxx}
\mathbf{R}_{\widetilde{\mathbf{x}}\widetilde{\mathbf{x}}}^{-1}&=&\widetilde{\mathbf{U}}_s\bm{\Sigma}^{-1}\widetilde{\mathbf{U}}_s^H~+~\mathsmaller{\frac{1}{\sigma_{w}^2}}\widetilde{\mathbf{U}}_w\widetilde{\mathbf{U}}_w^H.
\end{eqnarray}
To that end, let $\bm{\psi}$ and $\varphi$ be the two generic variables that run over all the possible values of $\bar{\bm{\psi}}_k$ and $\bar{\varphi}_k$, respectively. Then, right-multiplying (\ref{inverseRxx}) by $\widetilde{\mathbf{R}}^{(k)}_{ss}(\bm{\psi},\varphi)$ yields:
\begin{eqnarray}\label{invrela}
\!\!\!\!\!\!\!\!\!\!\!\!\!\mathbf{R}_{\widetilde{\mathbf{x}}\widetilde{\mathbf{x}}}^{-1}\widetilde{\mathbf{R}}^{(k)}_{ss}(\bm{\psi},\varphi)&\!\!=\!\!&\widetilde{\mathbf{U}}_s\bm{\Sigma}^{-1}\widetilde{\mathbf{U}}_s^H\widetilde{\mathbf{R}}^{(k)}_{ss}(\bm{\psi},\varphi)\nonumber\\
&\!\!\!\!&~~~~~~~~~~+~\mathsmaller{\frac{1}{\sigma_{w}^2}}\widetilde{\mathbf{U}}_w\widetilde{\mathbf{U}}_w^H\widetilde{\mathbf{R}}^{(k)}_{ss}(\bm{\psi},\varphi).
\end{eqnarray}
At relatively high SNR levels, the signal eigenvalues in $\bm{\Sigma}$ are relatively large and, therefore, the diagonal elements of $\bm{\Sigma}^{-1}$ are almost equal to zero. Consequently, the first term in the right-hand side of (\ref{invrela}) does not vary appreciably with $\bm{\psi}$ and $\varphi$.  The second term  in (\ref{invrela}) is thus dominant. Owing to (\ref{relsubs}), however, it is identically zero when $\bm{\psi}=\bar{\bm{\psi}}_k$  and $\varphi=\bar{\varphi}_k$ (for $k=1,2,\ldots, K$).
Therefore, at favorable SNR conditions, the quantity $\|\mathbf{R}_{\widetilde{\mathbf{x}}\widetilde{\mathbf{x}}}^{-1}~\!\widetilde{\mathbf{R}}^{(k)}_{ss}(\bm{\psi},\varphi)\|_{\textrm{Fro}}$ attains its minimum at $(\bar{\bm{\psi}}_k,~\bar{\varphi}_k)$ for each $k=1,2,\ldots, K$.   Based on this observation, the angular parameters can  be estimated jointly with the sources' noncircularity phases by resolving the following $K$ optimization problems:
\begin{eqnarray}\label{min_1}
\!\!\!\!\!\!\!\!\!\!\left[\widehat{\bar{\bm{\psi}}}_k,~\!\widehat{\bar{\varphi}}_k\right]&\!\!\!=\!\!\!&\underset{\bm{\psi},\bm{\varphi}}{\argmin}\left(\left\|\widehat{\mathbf{R}}_{\widetilde{\mathbf{x}}\widetilde{\mathbf{x}}}^{-1}\widetilde{\mathbf{R}}^{(k)}_{ss}(\bm{\psi},\varphi)\right\|_{\textrm{Fro}}^2\right)\!,\\
\label{min_2}&\!\!\!=\!\!\!&\underset{\bm{\psi},\bm{\varphi}}{\argmin}\left(\textrm{tr}\Big\{\widetilde{\mathbf{R}}^{(k)}_{ss}(\bm{\psi},\varphi)~\!\widehat{\mathbf{R}}_{\widetilde{\mathbf{x}}\widetilde{\mathbf{x}}}^{-2}~\widetilde{\mathbf{R}}^{(k)}_{ss}(\bm{\psi},\varphi)\Big\}\right)\!,
\end{eqnarray}
where $\widehat{\mathbf{R}}_{\widetilde{\mathbf{x}}\widetilde{\mathbf{x}}}$ is the sample-mean estimate of the actual extended covariance matrix, $\mathbf{R}_{\widetilde{\mathbf{x}}\widetilde{\mathbf{x}}}$, i.e.:
\begin{equation}
\widehat{\mathbf{R}}_{\widetilde{\mathbf{x}}\widetilde{\mathbf{x}}}~=~\frac{1}{N}\sum_{n=1} ^N\widetilde{\mathbf{x}}(n)\widetilde{\mathbf{x}}(n)^H,
\end{equation}
in which $N$ stands for the number of snapshots.
Further, if the sources have the same scatterers' angular distribution \big[i.e., $\widetilde{\mathbf{R}}^{(k)}_{ss}(\bm{\psi},\varphi) = \widetilde{\mathbf{R}}_{ss}(\bm{\psi},\varphi)$, $\forall k$\big],
then all the angular parameters can be estimated jointly by finding the location of the $K$ smallest  values of the common cost function:
\begin{eqnarray}\label{newcrit}
~\!\!\!f\big(\bm{\psi},\varphi\!~\big|\!~\widehat{\mathbf{R}}_{\widetilde{\mathbf{x}}\widetilde{\mathbf{x}}}^{-2}\big)&\triangleq&\textrm{tr}\Big\{\widetilde{\mathbf{R}}_{ss}(\bm{\psi},\varphi)~\!\widehat{\mathbf{R}}_{\widetilde{\mathbf{x}}\widetilde{\mathbf{x}}}^{-2}~\widetilde{\mathbf{R}}_{ss}(\bm{\psi},\varphi)\Big\},
\end{eqnarray}
where
\begin{eqnarray}
\label{common_cov}\!\!\widetilde{\mathbf{R}}_{ss}(\bm{\psi},\varphi)&=&\int\rho(\theta,\bm{\psi})~\!\widetilde{\mathbf{a}}(\theta,\varphi)~\!\widetilde{\mathbf{a}}(\theta,\varphi)^Hd\theta.
\end{eqnarray}
Note here that the cost function in (\ref{newcrit}) to be minimized requires a three-dimensional (3-D) search over the central DOA, $\Theta$, the angular spread, $\sigma$, and the noncircularity phase, $\varphi$. In the following, we will try to reduce the complexity of the proposed method by reducing the dimensionality of the cost function in (\ref{newcrit}).\\

Actually, using (\ref{extended-response}) in (\ref{common_cov}), it can be shown that:
\begin{eqnarray}
\label{extfm}\!\!\widetilde{\mathbf{R}}_{ss}(\bm{\psi},\varphi)
&=&\left(\begin{array}{cc}
\mathbf{R}_{ss}(\bm{\psi})&e^{j\varphi}\mathbf{R}'_{ss}(\bm{\psi})\\~\\
e^{-j\varphi}\mathbf{R}'^{*}_{ss}(\bm{\psi})&\mathbf{R}_{ss}^*(\bm{\psi})
\end{array}\right),
\end{eqnarray}
where $\mathbf{R}_{ss}(\bm{\psi})$ and $\mathbf{R}'_{ss}(\bm{\psi})$ are, respectively, the normalized \textit{conjugated} and \textit{unconjugated} noise-free auto-covariance matrices of the sources which are explicitly given by:
\begin{eqnarray}\label{conjcovmx}
\mathbf{R}_{ss}(\bm{\psi})&=&\int\rho(\theta,\bm{\psi})\mathbf{a}(\theta)\mathbf{a}(\theta)^Hd\theta,\\
\label{unccovmx}\mathbf{R}'_{ss}(\bm{\psi})&=&\int\rho(\theta,\bm{\psi})\mathbf{a}(\theta)\mathbf{a}(\theta)^Td\theta.
\end{eqnarray}
%
%
Assuming small angular spreads, we prove in Appendix A that $\mathbf{R}_{ss}(\bar{\bm{\psi}}_k)$ can be written for each $k$th source as:
\begin{eqnarray}
\!\!\!\!\!\!\!\!\label{newexpfm}\mathbf{R}_{ss}(\bar{\bm{\psi}}_k)&\simeq&\Big(\mathbf{a}(\bar{\Theta}_k)\!~\mathbf{a}(\bar{\Theta}_k)^H\Big)\varodot \mathbf{T}(\bar{\bm{\psi}}_k)\nonumber\\&\simeq&\bm{\Phi}(\bar{\Theta}_k)\!~\mathbf{T}(\bar{\bm{\psi}}_k)\!~\bm{\Phi}(\bar{\Theta}_k)^H,
\end{eqnarray}
with $\bm{\Phi}(\bar{\Theta}_k)~\triangleq~\textrm{diag}\{\mathbf{a}(\bar{\Theta}_k)\}$ and $\mathbf{T}(\bar{\bm{\psi}}_k)$ is a real-valued $(L\times L)$ symmetric matrix whose $(p,l)$th entry is given by:
\begin{eqnarray}\label{plelemfi}
[\mathbf{T}]_{pl}(\bar{\bm{\psi}}_k)&\!\!\!\!\!\!=\!\!\!\!\!\!\!\!&\int\!\!\!\rho_k(\theta,\bar{\bm{\psi}}_k)\!\cos\!\bigg(\!2\pi\! \left(f'_{p-1}(\bar{\Theta}_k)\!-\!f'_{l-1}(\bar{\Theta}_k)\right)\!(\theta\!-\!\bar{\Theta}_k)\!\bigg)\!d\theta,\nonumber\\
&\!\!\!\!\!\!\!\!\!\!\!\!\!\!&
\end{eqnarray}
and $f'_{p-1}(\theta)$ stands for the first derivative of $f_{p-1}(\theta)$ with respect to $\theta$.\\
In the same way, we also show  that the normalized \textit{unconjugated} noise-free covariance matrix, $\mathbf{R}'_{ss}(\bar{\bm{\psi}}_k)$, of  \textit{noncircular} ID sources can be be expressed as:
\begin{eqnarray}\label{newexpfmprim}
\mathbf{R}'_{ss}(\bar{\bm{\psi}}_k)&\!\!\!\simeq\!\!\!&\Big(\mathbf{a}(\bar{\Theta}_k)\!~\mathbf{a}(\bar{\Theta}_k)^T\Big)\varodot \mathbf{T}'(\bar{\bm{\psi}}_k)\nonumber\\
&\simeq&\bm{\Phi}(\bar{\Theta}_k)\!~\mathbf{T}'(\bar{\bm{\psi}}_k)\!~\bm{\Phi}(\bar{\Theta}_k)^T,
\end{eqnarray}
where $\mathbf{T}'(\bar{\bm{\psi}}_k)$ is also a real-valued $(L\times L)$ symmetric matrix whose $(p,l)$th entry is given by:
\begin{eqnarray}\label{plelemtprm}
[\mathbf{T}']_{pl}(\bar{\bm{\psi}}_k)&\!\!\!\!\!\!=\!\!\!\!\!\!\!\!&\int\!\!\!\rho_k(\theta,\bar{\bm{\psi}}_k)\!\cos\!\bigg(\!2\pi\!\left(f'_{p-1}(\bar{\Theta}_k)\!+\!f'_{l-1}(\bar{\Theta}_k)\right)\!(\theta\!-\!\bar{\Theta}_k)\!\bigg)\!d\theta.\nonumber\\
&\!\!\!\!\!\!\!\!\!\!\!\!\!\!&
\end{eqnarray}
Injecting (\ref{newexpfm}) and (\ref{newexpfmprim}) back into (\ref{extfm}) with the generic $\bm{\psi}$ and $\Theta$ being substituted for $\bar{\bm{\psi}}_k$ and $\bar{\Theta}_k$, respectively, and resorting to some straightforward  manipulations, it can be shown that:
\begin{eqnarray}
\widetilde{\mathbf{R}}_{ss}(\bm{\psi},\varphi)&\!\!=\!\!&\Big(\widetilde{\mathbf{a}}(\Theta,\varphi)\!~\widetilde{\mathbf{a}}(\Theta,\varphi)^H\Big)\varodot \widetilde{\mathbf{T}}(\bm{\psi}),\nonumber\\
\label{newexpextfm}&\!\!=\!\!&\widetilde{\bm{\Phi}}(\Theta,\varphi)~\!\widetilde{\mathbf{T}}(\bm{\psi})~\!\widetilde{\bm{\Phi}}(\Theta,\varphi)^H,
\end{eqnarray}
in which  $\widetilde{\bm{\Phi}}(\Theta,\varphi)~=~\textrm{diag}\{\widetilde{\mathbf{a}}(\Theta,\varphi)\}$ and
\begin{eqnarray}
\label{ext_T}\widetilde{\mathbf{T}}(\bm{\psi})&=&\left(\begin{array}{cc}
\mathbf{T}(\bm{\psi})&~~{\mathbf{T}'}(\bm{\psi})\\\\
{\mathbf{T}'}(\bm{\psi})&~~\mathbf{T}(\bm{\psi})
\end{array}\right).
\end{eqnarray}
For mathematical convenience, we also introduce the following notations:
\begin{eqnarray}
\label{matrix_A}\mathbf{A}(\bm{\psi})&\!\!=\!\!&\mathbf{T}^2(\bm{\psi})+{\mathbf{T}'}^2(\bm{\psi}),\\
\label{matrix_B}\mathbf{B}(\bm{\psi})&\!\!=\!\!&\mathbf{T}(\bm{\psi}){\mathbf{T}'}(\bm{\psi})+{\mathbf{T}'}(\bm{\psi})\mathbf{T}(\bm{\psi}),\\
\label{expR1}\widehat{\mathbf{R}}_1&\!\!=\!\!&\widehat{\mathbf{R}}_{\widetilde{\mathbf{x}}\widetilde{\mathbf{x}}}^{-2}(1:L,1:L),\\
\label{expR2}\widehat{\mathbf{R}}_2&\!\!=\!\!&\widehat{\mathbf{R}}_{\widetilde{\mathbf{x}}\widetilde{\mathbf{x}}}^{-2}(1:L,L+1:2L).
\end{eqnarray}
Then, plugging (\ref{newexpextfm}) back into (\ref{newcrit}), we prove after tedious manipulations (cf. Appendix B), that the angular parameters, $\{\bar{\bm{\psi}}_k\}_{k=1}^K$, can now be estimated by minimizing the following compressed cost function (i.e., that  depends on $\bm{\psi}$ only):
\begin{eqnarray}\label{expgpsi}
\!\!\!\!\!\!\!\!\!\!\!\!f_c\big(\bm{\psi}\!~\big|\!~\widehat{\mathbf{R}}_{\widetilde{\mathbf{x}}\widetilde{\mathbf{x}}}^{-2}\big)&\!\!\!\!&\nonumber\\&&\!\!\!\!\!\!\!\!\!\!\!\!\!\!\!\!\!\!\!\!\!\!\!\!=~\!\Re\bigg\{\textrm{tr}\Big\{\textrm{diag}\big\{\mathbf{a}(\Theta)\big\}\mathbf{A}(\bm{\psi})\textrm{diag}\big\{\mathbf{a}(\Theta)^H\big\}\widehat{\mathbf{R}}_1\Big\}\bigg\}\nonumber\\
&\!\!\!\!&\!\!\!\!\!\!-\bigg|~\!\textrm{tr}\Big\{\textrm{diag}\big\{\mathbf{a}(\Theta)\big\}\mathbf{B}(\bm{\psi})\textrm{diag}\big\{\mathbf{a}(\Theta)\big\}^T\widehat{\mathbf{R}}_2^*\Big\}\bigg|.
\end{eqnarray}
Note here that the first version of our proposed method defined by the cost function in (\ref{expgpsi}) is applicable for a general class of angular distributions (symmetric distributions with small angular spreads) and any planar array configuration. However, it requires the \textit{a priori} knowledge of the angular distributions to calculate the matrices $\mathbf{A}$ and $\mathbf{B}$ from (\ref{matrix_A}) and (\ref{matrix_B}), respectively.
Furthermore, finding the $K$ minima of (\ref{expgpsi}) with respect to $\bm{\psi}=\big[\Theta,~\sigma\big]^T$ still requires a two-dimensional (2-D) search over $\Theta$ and $\sigma$ and needs the angular distribution to be identical for all the sources to estimate jointly the angular parameters.
In the following, we will build upon some properties of the matrices $\mathbf{T}(\bm{\psi})$ and ${\mathbf{T}'}(\bm{\psi})$ in order to decouple the estimation of the central DOAs from that of the angular spreads [\ref{LEEE29}]. These properties are valid for any symmetric source's angular distribution with small angular spreads. Hence, the estimator can be implemented by two successive one-dimensional (1-D) parameter searches, thereby resulting in tremendous computational savings. Moreover, we will exploit these properties to establish unstructured models for $\mathbf{T}(\bm{\psi})$ and ${\mathbf{T}'}(\bm{\psi})$ that are totally oblivious to the symmetric sources' angular distributions. Therefore, we will obtain a new version of the proposed estimator that does not require the \textit{a priori} knowledge of the sources' angular distributions.

\section{Robust Version of the Proposed Estimator}\label{section_31}
To begin with, for any array configuration, recall that $\mathbf{T}(\bm{\psi})$ is a real-valued symmetric matrix whose expression is given by (\ref{plelemfi}).
Moreover, we prove in the following that if $f'_{p-1}(\bar{\Theta}_k)$ is expressed as follows\footnote{(\ref{simexpf}) means that the antenna array must be an equally-spaced linear array.}:
\begin{eqnarray}\label{simexpf}
f'_{p-1}(\bar{\Theta}_k)&=&(p-1)g(\bar{\Theta}_k),
\end{eqnarray}
where $g(\bar{\Theta}_k)$ is a transformation of the central DOA $\bar{\Theta}_k$, then $\mathbf{T}(\bar{\bm{\psi}}_k)$ is a symmetric Toeplitz matrix. In fact, injecting (\ref{simexpf}) in (\ref{plelemfi}), we show that $[\mathbf{T}]_{pl}(\bar{\bm{\psi}}_k)$ can be written as follows:
\begin{eqnarray}\label{simpexpT}
\!\!\!\![\mathbf{T}]_{pl}(\bar{\bm{\psi}}_k)&\!\!\!\!\!\!=\!\!\!\!\!\!&\int\! \rho_k(\theta,\bar{\bm{\psi}}_k)\cos\left(2\pi(p-l)g(\bar{\Theta}_k)\!(\theta\!-\!\bar{\Theta}_k)\right)\!d\theta.
\end{eqnarray}
From (\ref{simpexpT}), we can simply verify that:
\begin{eqnarray}
[\mathbf{T}]_{pl}(\bar{\bm{\psi}}_k)&\!\!\!=\!\!\!&[\mathbf{T}]_{(p+m)(l+m)}(\bar{\bm{\psi}}_k), \forall m.
\end{eqnarray}
Consequently, $\mathbf{T}(\bar{\bm{\psi}}_k)$ is a symmetric Toeplitz matrix and, therefore, it can be fully constructed from its first column vector denoted here as $\mathbf{t}_1$, i.e.:
\begin{eqnarray}\label{toeplitz}
\mathbf{T}(\bm{\psi})&=&\textrm{Toeplitz}\big(\mathbf{t}_1\big).
\end{eqnarray}
Moreover, for any symmetric angular distribution, we prove in Appendix C that if its angular spread verifies the following condition:
\begin{eqnarray}\label{condsig}
\sigma<\frac{1}{\sqrt{2}\pi(L-1)g(\bar{\Theta}_k)},
\end{eqnarray}
then the elements, $\{\mathbf{t}_1(l)\}_{l=1}^{L}$, of the vector,  $\mathbf{t}_1$, satisfy the following property:
\begin{equation}\label{constraint_1}
1~=~\mathbf{t}_1(1)~\geq~ \mathbf{t}_1(2)~\geq\ldots \geq~\mathbf{t}_1(L)~ \geq~ 0.
\end{equation}
(\ref{condsig}) is a nonrestrictive condition for propagation environments characterized by small angular spreads, e.g., macro-cell environments [\ref{LEEE350}-\ref{LEEE352}].
Actually, (\ref{constraint_1}) can be rewritten in the more succinct form\footnote{Note here that the notation $\mathbf{v}_1\leq \mathbf{v}_2$ for any tow $N-$dimensional vectors $\mathbf{x}=[x_1,x_2,\ldots, x_N]^T$ and $\mathbf{y}=[y_1,y_2,\ldots, y_N]^T$ means that $\mathbf{x}_n\leq y_n$ for $n=1,2,\ldots,N$.}:
\begin{eqnarray}\label{constraint_1_compact}
\mathbf{J}_L\mathbf{t}_1~\leq~\mathbf{e}_L,
\end{eqnarray}
where  $\mathbf{J}_n$ is  from now on a ($n\times n$) matrix given by:
\begin{eqnarray}
\mathbf{J}_n&=&\left(
\begin{array}{ccccc}
1&0&0&\cdots& 0\\
-1&1&0&\cdots& 0\\
0&-1&1&\ddots& 0\\
\vdots&\ddots&\ddots&\ddots&\vdots\\
0&\cdots&\cdots&-1&1\\
0&\cdots&\cdots&0&-1\end{array}\right),\nonumber
\end{eqnarray}
and $\mathbf{e}_n$ is a $n-$dimensional vector given by $\mathbf{e}_n=[1,0,\ldots,0]^T$.\\
For any array configuration, recall also that ${\mathbf{T}'}(\bm{\psi})$ is a real-valued symmetric matrix whose expression is given by (\ref{plelemtprm}). 
Moreover, if $f'_{p-1}(\bar{\Theta}_k)$ satisfies (\ref{simexpf}), we show that $[\mathbf{T}']_{pl}(\bar{\bm{\psi}}_k)$ can be written as:
\begin{eqnarray}\label{simpexpTpr}
\!\!\!\!\!\!\!\!\!\![\mathbf{T}']_{pl}(\bar{\bm{\psi}}_k)&\!\!\!\!\!\!=\!\!\!\!\!\!\!\!&\int\!\! \rho_k(\theta,\bar{\bm{\psi}}_k)\cos\!\left(2\pi(p+l-2)g(\bar{\Theta}_k)\!(\theta\!-\!\bar{\Theta}_k)\right)\!d\theta.
\end{eqnarray}
From (\ref{simpexpTpr}), we can see that ${\mathbf{T}'}(\bm{\psi})$ is a Hankel matrix. Therefore, it can be constructed from its first and last column vectors denoted, respectively,  as $\mathbf{t}'_1$ and $\mathbf{t}'_L$ as follows:
\begin{eqnarray}\label{hankel}
\mathbf{T}'(\bm{\psi})&=&\textrm{Hankel}\big(\mathbf{t}'_1,\mathbf{t}'_L\big).
\end{eqnarray}
Moreover, for any symmetric source's angular distribution, we also prove in Appendix C that if $\sigma<1/\big(2\sqrt{2}\pi(L-1)g(\bar{\Theta}_k)\big)$, then the elements of  $\mathbf{t}'_1$ and $\mathbf{t}'_L$  satisfy the following properties:
\begin{eqnarray}\label{constraint_2_compact}
\mathbf{J}_L\mathbf{t}'_1\leq\mathbf{e}_L,~~~
\mathbf{J}_L\mathbf{t}'_L\leq\mathbf{e}_L,~~~\textrm{and}~~~
\mathbf{t}'_1(L)=\mathbf{t}'_L(1).
\end{eqnarray}
Furthermore, we verify that the first column vector of ${\mathbf{T}'}(\bm{\psi})$ is identical to the first column vector of $\mathbf{T}(\bm{\psi})$, i.e., we have the following relation:
\begin{eqnarray}\label{relationt1t2}
 \mathbf{t}_1&=&\mathbf{t}'_1.
\end{eqnarray}
In order to exploit the interesting properties stated above in (\ref{constraint_1_compact}), (\ref{constraint_2_compact}) and (\ref{relationt1t2}), we consider an auxiliary vector $\mathbf{z}=[z_1,\ldots,z_{L-1},z_L,\ldots,z_{2L-2}]^T$ whose elements are all in $[0, 1]$ and sorted in decreasing order:
\begin{eqnarray}\label{rela_y}
1\geq z_1\geq z_2\geq \ldots \geq z_{L-1}\geq z_L\geq \ldots \geq z_{2L-2} \geq 0,
\end{eqnarray}
or equivalently:
\begin{eqnarray}\label{consminprobopt}
\mathbf{J}_{2L-1}\mathbf{z}\leq\mathbf{e}_{2L-1}.
\end{eqnarray}
Then, we  construct the following two auxiliary matrices:
\begin{eqnarray}
\mathbf{Z}&\!\!=\!\!&\textrm{Toeplitz}\Big(\big[1,\mathbf{z}(1:L-1)\big]\Big),\\
\mathbf{Z}'&\!\!=\!\!&\textrm{Hankel}\Big(\big[1,\mathbf{z}(1:L-1)\big],\mathbf{z}(L-1:2L-2)\Big),
\end{eqnarray}
which also verify the constraints in (\ref{constraint_1_compact}) and (\ref{constraint_2_compact}), respectively.
Therefore, bearing in mind the expressions of the matrices $\mathbf{A}$ and $\mathbf{B}$ in (\ref{matrix_A}) and (\ref{matrix_B}), respectively, it follows that instead of minimizing the 2-D criterion in (\ref{expgpsi}), one can start by solving the following 1-D constrained optimization problem in order to find the central DOAs:
\begin{eqnarray}\label{consminprob}
\!\!\!\!\!\!\!\!\!\widehat{\bar{\bm{\Theta}}}&\!\!=\!\!&\arg\underset{\Theta}{{\min}_K}\bigg(\underset{\mathbf{z}}{\min}~g(\Theta,\mathbf{z}) ~~~~\textrm{subject to}~~~~(\ref{consminprobopt})\bigg),
\end{eqnarray}
where
\begin{eqnarray}
\!\!\!\!g(\Theta,\mathbf{z})&\!\!\!=\!\!\!&\Re\Bigg\{\textrm{tr}\bigg\{\textrm{diag}\big\{\mathbf{a}(\Theta)\big\}\big(\mathbf{Z}^2+\mathbf{Z}'^2\big)\textrm{diag}\big\{\mathbf{a}(\Theta)^H\big\}\widehat{\mathbf{R}}_1\bigg\}\Bigg\}\nonumber\\
&\!\!\!\!\!\!&-~~\Bigg|~\!\textrm{tr}\bigg\{\textrm{diag}\big\{\mathbf{a}(\Theta)\big\}\big(\mathbf{Z}\mathbf{Z}'+\mathbf{Z}'\mathbf{Z}\big)\textrm{diag}\big\{\mathbf{a}(\Theta)^T\big\}\widehat{\mathbf{R}}_2^*\bigg\}\Bigg|.\nonumber\\
&\!\!\!\!\!\!&
\end{eqnarray}
The optimization task in (\ref{consminprob}) can be solved efficiently via the well-known \textit{sequential quadratic programming} (SQP) algorithm which is  a rapidly converging descent method for nonlinearly-constrained
optimization problems [\ref{sqp1}].
Interestingly enough, the estimator in (\ref{consminprob}) is also totally oblivious to the sources' angular distributions provided that the latter be symmetric. In fact, the auxiliary matrices $\mathbf{Z}$ and $\mathbf{Z}'$ involved in (\ref{consminprobopt}) were built for any symmetric angular distribution upon some general properties shared by $\mathbf{T}$ and $\mathbf{T}'$, respectively, and not their true expressions as required in (\ref{expgpsi}). Moreover, this estimator is applicable in the more challenging scenario where the sources
have different angular distributions. These are  actually quite precious degrees of freedom in practice since the  angular  distribution  may vary from one environment to another and/or from source to source in real-world scenarios.
After acquiring the central DOAs, $\widehat{\bar{\mathbf{\Theta}}}\!~=~\!\big[\widehat{\bar{\Theta}}_1,\widehat{\bar{\Theta}}_2,\ldots,\widehat{\bar{\Theta}}_K\big]$, as in (\ref{consminprob}), the angular spread pertaining to each $k$th source is estimated as follows:
\begin{eqnarray}\label{estiangsp1d}
\!\!\!\widehat{\bar\sigma}_k&=&\arg\underset{\sigma}{\min}~f_c\big(\widehat{\bar{\Theta}}_k,\sigma\!~\big|\!~\widehat{\mathbf{R}}_{\widetilde{\mathbf{x}}\widetilde{\mathbf{x}}}^{-2}\big),
\end{eqnarray}
where $f_c(.)$ is the compressed cost function already established in (\ref{expgpsi}).
Our estimator actually reduces to (\ref{consminprob}) and (\ref{estiangsp1d}), that is after going through three different derivation stages, from (\ref{newcrit}) to (\ref{expgpsi}), ultimately leading to our final robust solution.

\section{Statistical Properties}\label{section_4}
In order to assess the theoretical performance limits
of the proposed estimator, we will express its mean square error (MSE) analytically, based on the minimization of the original cost function, $f\big(\bm{\psi},\varphi~\!\big|\!~\widehat{\mathbf{R}}_{\widetilde{\mathbf{x}}\widetilde{\mathbf{x}}}^{-2}\big)$, given in (\ref{newcrit}) instead of the compressed one in (\ref{expgpsi}) due to the presence of the  nonlinear modulus operator in it.
 We will also use  $\bm{\alpha}_k=[\bar{\bm{\psi}}^T_k,\bar{\varphi}_k]^T=[\bar{\Theta}_k,\bar{\sigma}_k,\bar{\varphi}_k]^T$ to denote the entire unknown parameter vector pertaining to each $k$th  \textit{noncircular}  ID source. We further use:
\begin{eqnarray}\label{bias-definition}
\mathbf{b}(\widehat{\bm{\alpha}}_k)&\triangleq&\mathbb{E}\{\widehat{\bm{\alpha}}_k\}~-~\bm{\alpha}_k,
\end{eqnarray}
\begin{eqnarray}\label{variance_definition}
\mathbf{V}(\widehat{\bm{\alpha}}_k)&\triangleq&\mathbb{E}\Big\{\big(\widehat{\bm{\alpha}}_k-\mathbb{E}\{\widehat{\bm{\alpha}}_k\}\big)\big(\widehat{\bm{\alpha}}_k-\mathbb{E}\{\widehat{\bm{\alpha}}_k\}\big)^T\Big\},
\end{eqnarray}
to denote, respectively,  the bias vector and covariance matrix of
the estimate  $\widehat{\bm{\alpha}}_k$.  To begin with, it is easy to show that the mean square error (MSE), defined as $\textrm{MSE}(\widehat{\bm{\alpha}}_k)\triangleq\mathbb{E}\big\{(\widehat{\bm{\alpha}}_k-\bm{\alpha}_k)(\widehat{\bm{\alpha}}_k-\bm{\alpha}_k)^T\big\}$, is given by:
\begin{eqnarray}
\label{exp_mse}\textrm{MSE}(\widehat{\bm{\alpha}}_k)
&=&\mathbf{b}(\widehat{\bm{\alpha}}_k)\mathbf{b}^T(\widehat{\bm{\alpha}}_k)~+~\mathbf{V}(\widehat{\bm{\alpha}}_k).
\end{eqnarray}
Moreover, similar to [\ref{LEEE30}], let $\breve{\bm{\alpha}}_k$ denote the asymptotic estimate (obtained when the number of snapshots $N \rightarrow \infty$),  and define $\Delta\bm{\alpha}_k~\triangleq~\breve{\bm{\alpha}}_k~-~\bm{\alpha}_k$ and  $\Delta\breve{\bm{\alpha}}_k~\triangleq~\widehat{\bm{\alpha}}_k-\breve{\bm{\alpha}}_k$. Then, it immediately follows from (\ref{bias-definition}) that $\mathbf{b}(\widehat{\bm{\alpha}}_k)$  decomposes as the sum of the asymptotic bias and  the residual bias stemming from the finite-sample effects:
\begin{eqnarray}\label{expbias}
\mathbf{b}(\widehat{\bm{\alpha}}_k)&=&\mathbb{E}\{\Delta\bm{\alpha}_k\}~+~\mathbb{E}\{\Delta\breve{\bm{\alpha}}_k\}.
\end{eqnarray}
Furthermore, using some relatively straightforward algebraic manipulations, it can be shown that:
\begin{eqnarray}
\label{exp_cov}\mathbf{V}(\widehat{\bm{\alpha}}_k)&=&\mathbb{E}\big\{\Delta\breve{\bm{\alpha}}_k\Delta\breve{\bm{\alpha}}^T_k\big\}~-~\mathbb{E}\big\{\Delta\breve{\bm{\alpha}}_k\big\}\mathbb{E}\big\{\Delta\breve{\bm{\alpha}}^T_k\big\}.
\end{eqnarray}
Plugging (\ref{expbias}) and (\ref{exp_cov}) back into (\ref{exp_mse}), it follows that:
\begin{eqnarray}\label{MSE_0}
\!\!\!\!\!\!\!\!\!\!\!\!\!\!\!\!\textrm{MSE}(\widehat{\bm{\alpha}}_k)&\!\!=\!\!&\mathbb{E}\big\{\Delta\bm{\alpha}_k\Delta\bm{\alpha}^T_k\big\}~\!+~\!\mathbb{E}\big\{\Delta\bm{\alpha}_k\big\}\mathbb{E}\big\{\Delta\breve{\bm{\alpha}}_k\big\}^T\nonumber\\
&\!\!\!\!&~~~+~\!\mathbb{E}\big\{\Delta\breve{\bm{\alpha}}_k\big\}\mathbb{E}\big\{\Delta\bm{\alpha}_k\big\}^T+\mathbb{E}\big\{\Delta\breve{\bm{\alpha}}_k\Delta\breve{\bm{\alpha}}^T_k\big\}\!.
\end{eqnarray}
In order to establish an analytical expression for $\textrm{MSE}(\hat{\bm{\alpha}}_k)$, we will derive hereafter the four expectations involved in (\ref{MSE_0}) separately.  To do so, we use $\bm{\alpha}$ and $\mathbf{R}$ as generic variables for $\bm{\alpha}_k$ and $\widehat{\mathbf{R}}^{-2}_{\widetilde{\mathbf{x}}\widetilde{\mathbf{x}}}$, respectively. We also denote the gradient vector and Hessian matrix of the scalar-valued objective function, $f(\bm{\alpha}, \mathbf{R})$, as follows:
\begin{eqnarray}
\mathbf{f}(\bm{\alpha}|\mathbf{R})&\triangleq&\frac{\partial f(\bm{\alpha}|\mathbf{R})}{\partial\bm{\alpha}},~~~~~~~~\textrm{(gradient vector)}\nonumber\\
\mathbf{F}(\bm{\alpha}|\mathbf{R})&\triangleq&\frac{\partial^2 f(\bm{\alpha}|\mathbf{R})}{\partial\bm{\alpha}\partial\bm{\alpha}^T}. ~~~~~~~~\textrm{(Hessian matrix)}\nonumber
\end{eqnarray}
From (\ref{newcrit}), the $i^{th}$ element of the vector $\mathbf{f}(\bm{\alpha}|\mathbf{R})$ is given by:
\begin{eqnarray}\label{gradient_definition}
\big[\mathbf{f}(\bm{\alpha}|\mathbf{R})\big]_i&\triangleq&\frac{\partial f(\bm{\alpha}|\mathbf{R})}{\partial\alpha_i}~=~\textrm{tr}\left\{\mathbf{R}~\!\widetilde{\mathbf{R}}^{[i]}_{ss}\right\},
\end{eqnarray}
where
\begin{eqnarray}\label{R_i}
\widetilde{\mathbf{R}}^{[i]}_{ss}&\triangleq&\widetilde{\mathbf{R}}_{ss}\frac{\partial\widetilde{\mathbf{R}}_{ss}}{\partial\alpha_i}~+~\frac{\partial\widetilde{\mathbf{R}}_{ss}}{\partial\alpha_i}\widetilde{\mathbf{R}}_{ss}.
\end{eqnarray}
Furthermore, the  entries of the Hessian matrix, $\mathbf{F}(\bm{\alpha}|\mathbf{R})$, are  obtained as follows:
\begin{eqnarray}\label{Hessian_definition}
\big[\mathbf{F}(\bm{\alpha}|\mathbf{R})\big]_{ij}&\triangleq&\frac{\partial^{2}f(\bm{\alpha}|\mathbf{R})}{\partial\alpha_i\partial\alpha_j}~=~\textrm{tr}\left\{\mathbf{R}~\!\widetilde{\mathbf{R}}^{[i,j]}_{ss}\right\}.
\end{eqnarray}
where
\begin{eqnarray}
\!\!\!\widetilde{\mathbf{R}}^{[i,j]}_{ss}&&\nonumber\\
&&\!\!\!\!\!\!\!\!\!\!\!\!\!\!\!\!\!\!\triangleq\frac{\partial\widetilde{\mathbf{R}}_{ss}}{\partial\alpha_i}\frac{\partial\widetilde{\mathbf{R}}_{ss}}{\partial\alpha_j}+\widetilde{\mathbf{R}}_{ss}\frac{\partial^{2}\widetilde{\mathbf{R}}_{ss}}{\partial\alpha_i\partial\alpha_j}+\frac{\partial^{2}\widetilde{\mathbf{R}}_{ss}}{\partial\alpha_i\partial\alpha_j}\widetilde{\mathbf{R}}_{ss}+\frac{\partial\widetilde{\mathbf{R}}_{ss}}{\partial\alpha_j}\frac{\partial\widetilde{\mathbf{R}}_{ss}}{\partial\alpha_i}\nonumber
\end{eqnarray}

\subsection{\textbf{Derivation of} ~$\mathbb{E}\{\Delta\bm{\alpha}_k\}$ ~\textbf{and}~ $\mathbb{E}\{\Delta\bm{\alpha}_k\Delta\bm{\alpha}_k^T\}$:}\label{subsection_1}
First, using the properties of the complex Wishart distribution [\ref{LEEE35}, p. $273$], it can be shown that the asymptotic sample-mean estimate of the extended covariance matrix,  $\breve{\mathbf{R}}_{\widetilde{\mathbf{x}}\widetilde{\mathbf{x}}}=\lim_{N\to\infty} \widehat{\mathbf{R}}_{\widetilde{\mathbf{x}}\widetilde{\mathbf{x}}}$,  is a consistent estimate of the Hermitian matrix $\mathbf{R}_{\widetilde{\mathbf{x}}\widetilde{\mathbf{x}}}$. This means that as $N \rightarrow \infty$, we have:
\begin{eqnarray}\label{relainvinvR}
\breve{\mathbf{R}}_{\widetilde{\mathbf{x}}\widetilde{\mathbf{x}}}^{-2}&=&\mathbf{R}_{\widetilde{\mathbf{x}}\widetilde{\mathbf{x}}}^{-2}.
\end{eqnarray}
To derive the asymptotic bias, $\Delta\bm{\alpha}_k$, we use as in [\ref{LEEE36}, \ref{LEEE37}] the first-order Taylor series expansion of $\mathbf{f}(\bm{\alpha},\mathbf{R}_{\widetilde{\mathbf{x}}\widetilde{\mathbf{x}}}^{-2})$  around the actual parameter vector $\bm{\alpha}_k$:
\begin{eqnarray}\label{gradient_1}
\!\!\!\!\!\!\!\!\mathbf{f}\left(\bm{\alpha}\big|\mathbf{R}_{\widetilde{\mathbf{x}}\widetilde{\mathbf{x}}}^{-2}\right)&\!\simeq\!&\mathbf{f}\left(\bm{\alpha}_k\big|\mathbf{R}_{\widetilde{\mathbf{x}}\widetilde{\mathbf{x}}}^{-2}\right)\!~+\!~\mathbf{F}\left(\bm{\alpha}_k\big|\mathbf{R}_{\widetilde{\mathbf{x}}\widetilde{\mathbf{x}}}^{-2}\right)\big(\bm{\alpha}-\bm{\alpha}_k\big).
\end{eqnarray}
By noticing that the asymptotic estimate, $\breve{\bm{\alpha}}_k$, also minimizes $f(\bm{\alpha}|\mathbf{R}_{\widetilde{\mathbf{x}}\widetilde{\mathbf{x}}}^{-2})$, it follows that $\mathbf{f}(\breve{\bm{\alpha}}_k|\mathbf{R}_{\widetilde{\mathbf{x}}\widetilde{\mathbf{x}}}^{-2})=\mathbf{0}$. Therefore, by evaluating (\ref{gradient_1}) at  $\bm{\alpha}=\breve{\bm{\alpha}}_k$,  it follows that:
\begin{eqnarray}\label{gradient_1_1}
\mathbf{f}\left(\bm{\alpha}_k\big|\mathbf{R}_{\widetilde{\mathbf{x}}\widetilde{\mathbf{x}}}^{-2}\right)~+~\mathbf{F}\left(\bm{\alpha}_k\big|\mathbf{R}_{\widetilde{\mathbf{x}}\widetilde{\mathbf{x}}}^{-2}\right)(\breve{\bm{\alpha}}_k-\bm{\alpha}_k)&\simeq&\bm{0},
\end{eqnarray}
from which $\Delta\bm{\alpha}_k \triangleq (\breve{\bm{\alpha}}_k-\bm{\alpha}_k)$ is obtained as:
 \begin{eqnarray}
\Delta\bm{\alpha}_k&\simeq&-~\mathbf{F}^{-1}\left(\bm{\alpha}_k\big|\mathbf{R}_{\widetilde{\mathbf{x}}\widetilde{\mathbf{x}}}^{-2}\right)\mathbf{f}\left(\bm{\alpha}_k\big|\mathbf{R}_{\widetilde{\mathbf{x}}\widetilde{\mathbf{x}}}^{-2}\right).
\end{eqnarray}
Consequently, the approximate expression for the asymptotic bias, $\Delta\bm{\alpha}_k$, and $\mathbb{E}\big\{\Delta\bm{\alpha}_k\Delta\bm{\alpha}^T_k\big\}$ are obtained as follows:
\begin{eqnarray}
\mathbb{E}\{\Delta\bm{\alpha}_k\}&\simeq&-~\mathbf{F}^{-1}\left(\bm{\alpha}_k\big|\mathbf{R}_{\widetilde{\mathbf{x}}\widetilde{\mathbf{x}}}^{-2}\right)\mathbf{f}\left(\bm{\alpha}_k\big|\mathbf{R}_{\widetilde{\mathbf{x}}\widetilde{\mathbf{x}}}^{-2}\right),\nonumber
\end{eqnarray}
and
\begin{eqnarray}
\mathbb{E}\{\Delta\bm{\alpha}_k\Delta\bm{\alpha}_k^T\}&&\nonumber\\
&&\!\!\!\!\!\!\!\!\!\!\!\!\!\!\!\!\!\!\!\!\!\!\!\!\!\!\!\!\!\!\!\!\!\!\!\!\!\!\simeq\mathbf{F}^{-1}\left(\bm{\alpha}_k\big|\mathbf{R}_{\widetilde{\mathbf{x}}\widetilde{\mathbf{x}}}^{-2}\right)\mathbf{f}\left(\bm{\alpha}_k\big|\mathbf{R}_{\widetilde{\mathbf{x}}\widetilde{\mathbf{x}}}^{-2}\right)\mathbf{f}\left(\bm{\alpha}_k\big|\mathbf{R}_{\widetilde{\mathbf{x}}\widetilde{\mathbf{x}}}^{-2}\right)^T\mathbf{F}^{-1}\left(\bm{\alpha}_k\big|\mathbf{R}_{\widetilde{\mathbf{x}}\widetilde{\mathbf{x}}}^{-2}\right). \nonumber
\end{eqnarray}
\subsection{\textbf{Derivation of} ~$\mathbb{E}\{\Delta\breve{\bm{\alpha}}_k\}$ ~\textbf{and}~ $\mathbb{E}\{\Delta\breve{\bm{\alpha}}_k\Delta\breve{\bm{\alpha}}_k^T\}$:}\label{subsection_2}
After tedious algebraic  manipulations, we also show in Appendix D that $\Delta\breve{\bm{\alpha}}_k$ is expressed as follows:
\begin{eqnarray}\label{additional_bias_00}
\Delta\breve{\bm{\alpha}}_k&=&\mathbf{F}^{-1}(\breve{\bm{\alpha}}_k|\mathbf{R}_{\widetilde{\mathbf{x}}\widetilde{\mathbf{x}}}^{-2})~\!\mathbf{v}(\breve{\bm{\alpha}}_k|\mathbf{R}_{\widetilde{\mathbf{x}}\widetilde{\mathbf{x}}}^{-2},\widehat{\mathbf{R}}_{\widetilde{\mathbf{x}}\widetilde{\mathbf{x}}}^{-2}),
\end{eqnarray}
where $\mathbf{v}(\breve{\bm{\alpha}}_k|\mathbf{R}_{\widetilde{\mathbf{x}}\widetilde{\mathbf{x}}}^{-2},\widehat{\mathbf{R}}_{\widetilde{\mathbf{x}}\widetilde{\mathbf{x}}}^{-2})$ is a 3-dimensional vector whose $i^{th}$ element is explicitly given by
\begin{eqnarray}
\!\!\!\!\!\!\!\!\!\!\!\!\!\!\!v_i\big(\breve{\bm{\alpha}}_k|\mathbf{R}_{\widetilde{\mathbf{x}}\widetilde{\mathbf{x}}}^{-2},\widehat{\mathbf{R}}_{\widetilde{\mathbf{x}}\widetilde{\mathbf{x}}}^{-2}\big)&\!\!\!\!\!\!&\nonumber \\
&&\!\!\!\!\!\!\!\!\!\!\!\!\!\!\!\!\!\!\!\!\!\!\!\!=\!~\textrm{tr}\left\{\left[\frac{\partial}{\partial\mathbf{R}}\textrm{tr}\left\{\mathbf{R}~\!\widetilde{\mathbf{R}}^{[i]}_{ss}\right\}\right]^T\!\!\!\Delta\mathbf{R}_{\widetilde{\mathbf{x}}\widetilde{\mathbf{x}}}^{-2}\right\}\Bigg|_{\substack{\!\!\bm{\alpha}\!~=\!~\breve{\bm{\alpha}}_k\\~\mathbf{R}\!~=\!~\mathbf{R}_{\widetilde{\mathbf{x}}\widetilde{\mathbf{x}}}^{-2}}}\!\!\!.
\end{eqnarray}
Recall here that $\widetilde{\mathbf{R}}^{[i]}_{ss}$ was already defined in (\ref{R_i}) and we further define
$\Delta\mathbf{R}_{\widetilde{\mathbf{x}}\widetilde{\mathbf{x}}}^{-2}$ as follows:
\begin{eqnarray}\label{DeltaR}
\Delta\mathbf{R}_{\widetilde{\mathbf{x}}\widetilde{\mathbf{x}}}^{-2}&\triangleq&\widehat{\mathbf{R}}_{\widetilde{\mathbf{x}}\widetilde{\mathbf{x}}}^{-2}~-~\mathbf{R}_{\widetilde{\mathbf{x}}\widetilde{\mathbf{x}}}^{-2}.
\end{eqnarray}
 Then, by exploiting the fact that $\partial\textrm{tr}\{\mathbf{A}\mathbf{B}\}/\partial \mathbf{A}=\mathbf{B}^T$ for any two matrices $\mathbf{A}$ and $\mathbf{B}$, it follows that:
\begin{eqnarray}\label{v_i_detailed}
\!\!\!\!\!\!\!\!\!\!\!\!\!\!\!\!\!\!\!\!v_i\big(\breve{\bm{\alpha}}_k|\mathbf{R}_{\widetilde{\mathbf{x}}\widetilde{\mathbf{x}}}^{-2},\widehat{\mathbf{R}}_{\widetilde{\mathbf{x}}\widetilde{\mathbf{x}}}^{-2}\big)&=&
\textrm{tr}\left\{\!\widetilde{\mathbf{R}}^{[i]}_{ss}~\!\Delta\mathbf{R}_{\widetilde{\mathbf{x}}\widetilde{\mathbf{x}}}^{-2}\right\}\nonumber\\
&=&\left[\textrm{vec}\big\{\widetilde{\mathbf{R}}^{[i]T}_{ss}\big\}\right]^T\textrm{vec}\big\{\Delta\mathbf{R}_{\widetilde{\mathbf{x}}\widetilde{\mathbf{x}}}^{-2}\big\},
\end{eqnarray}
where the last equality follows from the identity $\textrm{tr}\{\mathbf{A}\mathbf{B}\}=\textrm{vec}^T\{\mathbf{A}^T\}\textrm{vec}\{\mathbf{B}\}$. Consequently, the vector $\mathbf{v}\big(\breve{\bm{\alpha}}_k|\mathbf{R}_{\widetilde{\mathbf{x}}\widetilde{\mathbf{x}}}^{-2},\widehat{\mathbf{R}}_{\widetilde{\mathbf{x}}\widetilde{\mathbf{x}}}^{-2}\big)$ is expressed as follows:
\begin{eqnarray}\label{vector_v}
\mathbf{v}\big(\breve{\bm{\alpha}}_k|\mathbf{R}_{\widetilde{\mathbf{x}}\widetilde{\mathbf{x}}}^{-2},\widehat{\mathbf{R}}_{\widetilde{\mathbf{x}}\widetilde{\mathbf{x}}}^{-2}\big)&=& \mathbf{G}_{ss}^T\textrm{vec}\big\{\Delta\mathbf{R}_{\widetilde{\mathbf{x}}\widetilde{\mathbf{x}}}^{-2}\big\},
\end{eqnarray}
where the matrix $\mathbf{G}_{ss}$ is given by
\begin{eqnarray}
\!\!\!\!\!\!\!\!\!\mathbf{G}_{ss}&=& \Big[\textrm{vec}\big\{\widetilde{\mathbf{R}}^{(1)T}_{ss}\big\}~~\textrm{vec}\big\{\widetilde{\mathbf{R}}^{(2)T}_{ss}\big\}~~\textrm{vec}\big\{\widetilde{\mathbf{R}}^{(3)T}_{ss}\big\}\Big].
 \end{eqnarray}
 Plugging (\ref{vector_v}) back into (\ref{additional_bias_00}), one obtains:
 \begin{eqnarray}\label{additional_bias_2}
\Delta\breve{\bm{\alpha}}_k&=&\mathbf{F}^{-1}(\breve{\bm{\alpha}}_k|\mathbf{R}_{\widetilde{\mathbf{x}}\widetilde{\mathbf{x}}}^{-2})~\mathbf{G}_{ss}^T\textrm{vec}\big\{\Delta\mathbf{R}_{\widetilde{\mathbf{x}}\widetilde{\mathbf{x}}}^{-2}\big\},
\end{eqnarray}
 whose expectation yields the required residual bias as follows:
 \begin{eqnarray}\label{additional_bias1}
 \!\!\!\!\!\!\!\!\!\!\!\!\mathbb{E}\big\{\Delta\breve{\bm{\alpha}}_k\big\}&\!\!=\!\!& \mathbf{F}^{-1}\left(\breve{\bm{\alpha}}_k|\mathbf{R}_{\widetilde{\mathbf{x}}\widetilde{\mathbf{x}}}^{-2}\right)~\!\mathbf{G}_{ss}^T ~\!\textrm{vec}\Big\{\mathbb{E}\big\{\Delta\mathbf{R}_{\widetilde{\mathbf{x}}\widetilde{\mathbf{x}}}^{-2}\big\}\Big\}.
\end{eqnarray}
Furthermore, in presence of noncircular signals,  it can be shown  that $\mathbb{E}\{\Delta\mathbf{R}_{\widetilde{\mathbf{x}}\widetilde{\mathbf{x}}}^{-2}\}$ is accurately approximated by\footnote{See [\ref{LEEE30}] and [\ref{LEEE37}] for more details about the proof in the case of \textit{circular} sources that we generalize here to the  \textit{noncircular} case using the appropriate extended covariance matrices.}:
\begin{eqnarray}
\mathbb{E}\Big\{\Delta\mathbf{R}_{\widetilde{\mathbf{x}}\widetilde{\mathbf{x}}}^{-2}\Big\}&\simeq&\frac{1}{N-2L}\sum_{n=1}^{2L}\frac{1}{\lambda_n^2}\widetilde{\bm{e}}_n\widetilde{\bm{e}}^{H}_n,\nonumber
\end{eqnarray}
where $\widetilde{\mathbf{e}}_n$ is an eigenvector associated to  the $n^{th}$ eigenvalue, $\lambda_n$, of the extended covariance matrix $\mathbf{R}_{\widetilde{\mathbf{x}}\widetilde{\mathbf{x}}}$. From (\ref{additional_bias_2}), it also immediately follows that:
\begin{eqnarray}\label{appexpcovmx}
\!\!\!\!\!\!\!\!\!\!\!\!\mathbb{E}\Big\{\Delta\breve{\bm{\alpha}}_k\Delta\breve{\bm{\alpha}}_k^T\Big\}&&\nonumber\\
&&\!\!\!\!\!\!\!\!\!\!\!\!\!\!\!\!\!\!\!\!\!\!\!\!=~\mathbf{F}^{-1}\big(\breve{\bm{\alpha}}_k~\!|~\!\mathbf{R}_{\widetilde{\mathbf{x}}\widetilde{\mathbf{x}}}^{-2}\big)~\!\mathbf{G}_{ss}^T~\!\mathbf{H}\!~\mathbf{G}_{ss}~\!\mathbf{F}^{-1}\big(\breve{\bm{\alpha}}_k~\!|~\!\mathbf{R}_{\widetilde{\mathbf{x}}\widetilde{\mathbf{x}}}^{-2}\big),
\end{eqnarray}
where
\begin{eqnarray}
\!\!\!\!\!\!\!\!\!\mathbf{H}&=&\mathbb{E}\Big\{\textrm{vec}\big\{\Delta\mathbf{R}_{\widetilde{\mathbf{x}}\widetilde{\mathbf{x}}}^{-2}\big\}\textrm{vec}\big\{\Delta\mathbf{R}_{\widetilde{\mathbf{x}}\widetilde{\mathbf{x}}}^{-2}\big\}^T\Big\}.
\end{eqnarray}
The  entries of $\mathbf{H}$ are also evaluated using the following accurate approximation:
\begin{eqnarray}\label{covdeltaR}
\!\!\!\!\!\!\!\!\!\!\mathbb{E}\Big\{\left[\Delta\mathbf{R}_{\widetilde{\mathbf{x}}\widetilde{\mathbf{x}}}^{-2}\right]_{ij}\left[\Delta\mathbf{R}_{\widetilde{\mathbf{x}}\widetilde{\mathbf{x}}}^{-2}\right]_{pl}\Big\}&&\nonumber\\&&
\!\!\!\!\!\!\!\!\!\!\!\!\!\!\!\!\!\!\!\!\!\!\!\!\!\!\!\!\!\!\!\!\!\!\!\!\!\!\!\!\!\!\!\!\!\!\!\!\!\!\!\!\!\!\!\!\simeq~\frac{1}{N-2L}\sum_{n=1}^{2L}\sum_{n'=1}^{2L}\!\!\omega_{nn'}\left[\widetilde{\mathbf{e}}_n\widetilde{\mathbf{e}}^{H}_n\right]_{il}\left[\widetilde{\mathbf{e}}_{n'}\widetilde{\mathbf{e}}^{H}_{n'}\right]_{pj},
\end{eqnarray}
in which the weighting coefficients, $\omega_{nn'}$, are simply given by:
\begin{eqnarray}
\omega_{nn'}&=&\lambda_n^{-1}\lambda_{n'}^{-1}\big(\lambda_n^{-1}+\lambda_{n'}^{-1}\big)^2.
\end{eqnarray}
\section{New CRLB for Noncircular Gaussian Distributed Signals Generated from ID Sources}\label{section_5}
In this section, we assume that the transmitted signals $\{\bm{s}(t)\}_{t=1, 2,\ldots, N}$ are zero-mean Gaussian distributed and generated from \textit{noncircular} ID sources. We also assume that the noncircularity rate of the signals is $0\leq\gamma\leq 1$. Now recall from (\ref{exextcomxRxx}) that the extended covariance matrix of the received signals is given by:
\begin{eqnarray}
\mathbf{R}_{\widetilde{\mathbf{x}}\widetilde{\mathbf{x}}}&\!\!=\!\!&\left(
\begin{array}{cc}
\mathbf{R}_{\mathbf{x}\mathbf{x}}~~~~~~\mathbf{R}^{'}_{\mathbf{x}\mathbf{x}}\\
{\mathbf{R}'}_{\mathbf{x}\mathbf{x}}^*~~~~~~\mathbf{R}_{\mathbf{x}\mathbf{x}}^*
\end{array}\right).
\end{eqnarray}
Moreover, using (\ref{conjcovmx}) and (\ref{unccovmx}) in (\ref{matrx}) and (\ref{secondmatrx}), respectively, it follows that:
\begin{eqnarray}
\label{matrx_CRB}\!\!\!\!\!\!\!\!\!\!\!\mathbf{R}_{\mathbf{x}\mathbf{x}}&\!\!=\!\!&\displaystyle\sum_{k=1}^{K}\sigma_{s_k}^2\mathbf{R}^{(k)}_{ss}(\bar{\bm{\psi}}_k)~+~\sigma_{w}^2\mathbf{I}_L,\\
\label{secondmatrx_CRB}\!\!\!\!\!\!\!\!\!\!\!\mathbf{R}'_{\mathbf{x}\mathbf{x}}&\!\!=\!\!&\displaystyle\sum_{k=1}^{K}\sigma_{s_k}^2\mathbf{R'}^{(k)}_{ss}(\bar{\bm{\psi}}_k).
\end{eqnarray}
Then, using (\ref{newexpfm}) and (\ref{newexpfmprim}) in (\ref{matrx_CRB}) and (\ref{secondmatrx_CRB}), respectively, leads to:
\begin{eqnarray}
\!\!\!\!\!\!\!\!\!\!\!\label{conj_cov_CRB}\mathbf{R}_{\mathbf{x}\mathbf{x}}&=&\displaystyle\sum_{k=1}^{K}\sigma_{s_k}^2\bm{\Phi}(\bar{\Theta}_k)\mathbf{T}(\bar{\bm{\psi}}_k)\bm{\Phi}(\bar{\Theta}_k)^H~+~\sigma_{w}^2\mathbf{I}_L,\\
\!\!\!\!\!\!\!\!\!\!\!\label{unconj_cov_CRB}\mathbf{R}'_{\mathbf{x}\mathbf{x}}&=&\displaystyle\sum_{k=1}^{K}\sigma_{s_k}^2
e^{j\bar{\varphi}_k}\gamma_k\bm{\Phi}(\bar{\Theta}_k){\mathbf{T}'}(\bar{\bm{\psi}}_k)\bm{\Phi}(\bar{\Theta}_k)^T.
\end{eqnarray}
Recall also that the explicit expressions of $\bm{\Phi}(\bar{\Theta}_k)$, $\mathbf{T}(\bar{\bm{\psi}}_k)$ and ${\mathbf{T}'}(\bar{\bm{\psi}}_k)$ were already given in Section \ref{section_3}. Our goal in this section is to find the CRLB of the unknown parameters of interest (i.e., namely the angular parameters) which are gathered in the following vector:
\begin{eqnarray}\label{interestparams}
\bm{\eta}&\triangleq&\left[\bm{\Theta}^T,~\bm{\sigma}^T\right]^T.
\end{eqnarray}
 The unknown nuisance parameters which are the noise variance, $\sigma_w$, the sources' powers, $\bm{\beta}~\triangleq~\left[\sigma_{s_1}^2,\ldots,\sigma_{s_K}^2\right]^T$, and their noncircularity phases, $\bm{\varphi}~\triangleq~\left[\varphi_1,\ldots,\varphi_K\right]^T$, are also gathered in the vector:
\begin{eqnarray}\label{nuisanceparams}
\bm{\xi}&\triangleq&\left[\bm{\beta}^T,~\bm{\varphi}^T,~\!\sigma_w^2\right]^T.
\end{eqnarray}
We will also group all the parameters in (\ref{nuisanceparams}) an (\ref{interestparams}) in a single vector:
\begin{eqnarray}
\label{wholpara}\bm{\upsilon}&\triangleq&\left[\bm{\eta}^T,~\bm{\xi}^T\right]^T.
\end{eqnarray}
The CRLB of the entire unknown parameter vector, $\bm{\upsilon}$, is defined as follows [\ref{LEEE349}]:
\begin{eqnarray}
\textrm{CRLB}(\bm{\upsilon})&\triangleq&\mathbf{I}^{-1}(\bm{\upsilon}),
\end{eqnarray}
where $\mathbf{I}(\bm{\upsilon})$ is the so-called  Fisher information matrix (FIM).
Since the extended snapshot vectors, $\{\widetilde{\mathbf{x}}(t)\}_{t=1}^N$, defined in (\ref{extrecvect}) are mutually independent, then according to [\ref{LEEE38}] the $(i,j)$th entry of the FIM associated to the underlying estimation problem is given by:
\begin{eqnarray}
\left[\mathbf{I}\right]_{ij}&=&\mathsmaller{\frac{N}{2}}\textrm{tr}\left\{\frac{\partial\mathbf{R}_{\widetilde{\mathbf{x}}\widetilde{\mathbf{x}}}}{\partial\upsilon_i}\mathbf{R}_{\widetilde{\mathbf{x}}\widetilde{\mathbf{x}}}^{-1}\frac{\partial\mathbf{R}_{\widetilde{\mathbf{x}}\widetilde{\mathbf{x}}}}{\partial\upsilon_j}\mathbf{R}_{\widetilde{\mathbf{x}}\widetilde{\mathbf{x}}}^{-1}\right\},
\end{eqnarray}
where $\upsilon_i$ is the $i$th element of the whole parameter vector given in (\ref{wholpara}).
Using (\ref{conj_cov_CRB}) and (\ref{unconj_cov_CRB}), we show in Appendix E that the CRLB for the angular parameters alone in presence of uncorrelated ID \textit{noncircular} sources is explicitly given by:
\begin{eqnarray}
\textrm{CRLB}(\bm{\eta})&=&\left(\mathbf{I}_{\bm{\eta},\bm{\eta}}-\mathbf{I}_{\bm{\xi},\bm{\eta}}^T\mathbf{I}_{\bm{\xi},\bm{\xi}}^{-1}\mathbf{I}_{\bm{\xi},\bm{\eta}}\right)^{-1},
\end{eqnarray}
where the expressions of $\mathbf{I}_{\bm{\eta},\bm{\eta}}$, $\mathbf{I}_{\bm{\xi},\bm{\eta}}$, and $\mathbf{I}_{\bm{\xi},\bm{\xi}}$ are provided in Appendix E.

\section{Simulation Results}\label{section_6}
In this section, we assess the performance  of the newly proposed method
and gauge it against the most recent state-of-the-art techniques that are geared toward multiple ID sources, namely ESB [\ref{LEEE30}] and RGC  [\ref{LEEE29}]. Although the latter were derived specifically for ID \textit{circular} sources, they can be applied to the \textit{noncircular} case as well after completely ignoring the non-zero \textit{unconjugated} covariance matrix. All the methods will be also gauged against the CRLB.
In all simulations, we consider complex Gaussian transmitted signals and a uniform linear array of $6$ sensors separated by half a wavelength.

\subsection{Assessment of the new estimator}\label{subsection_6_1}

\begin{figure}[!ht]
\vskip -0.2 cm
\begin{centering}
\includegraphics[scale=0.53]{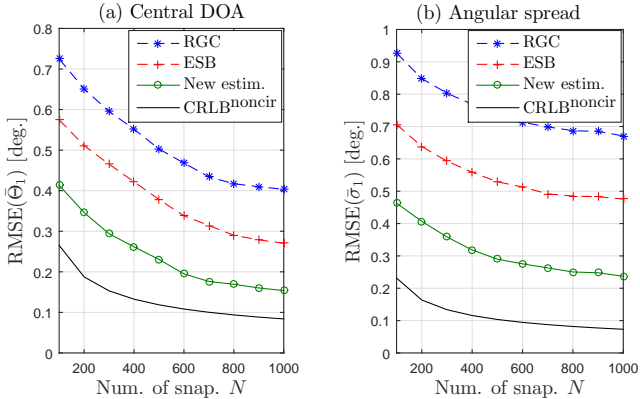}
\caption{RMSE of the three estimators versus $N$ for SNR $=5$ dB.}
\label{rmsesnap}
\end{centering}
\end{figure}

\begin{figure}[!ht]
\vskip -0.3 cm\hskip -0.3cm
\includegraphics[scale=0.53]{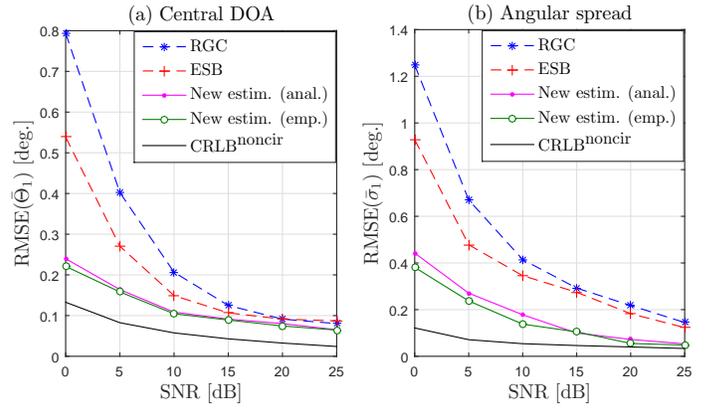}
\vskip -0.1cm
\caption{RMSE of the three  estimators versus SNR for $N=1000$, sources with the same angular distribution.}
\label{rmsett1ang}
\end{figure}

\begin{figure}[!ht]
\vskip -0.2 cm\hskip -0.15cm
\begin{centering}
\includegraphics[scale=0.53]{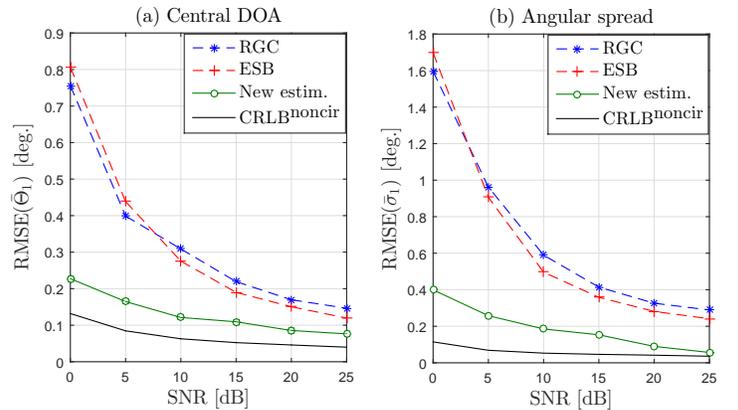}
\vskip -0.1cm
\caption{RMSE of the three  estimators versus SNR for $N=1000$, sources with different angular distributions.}
\label{rmset1angexp2}
\end{centering}
\end{figure}

\begin{figure}[!ht]
\vskip -0.2 cm\hskip -0.3cm
\begin{centering}
\includegraphics[scale=0.53]{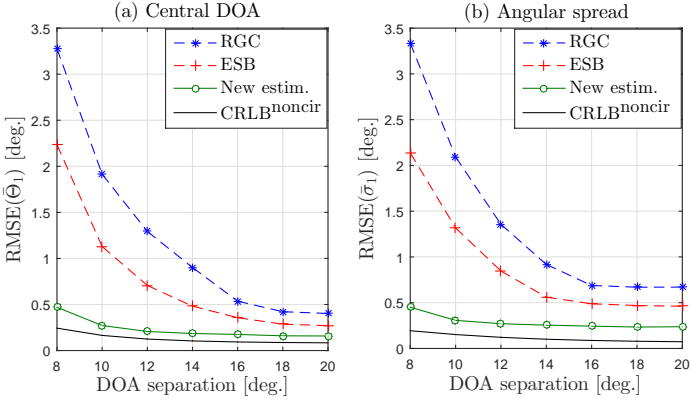}
\vskip -0.1cm
\caption{RMSE of the three  estimators versus DOA separation for $N=1000$ and  SNR $=5$ dB, sources with the same angular distributions.}
\label{rmset1angexp3}
\end{centering}
\end{figure}

\begin{figure}[!ht]
\vskip -0.2 cm
\begin{centering}
\includegraphics[scale=0.53]{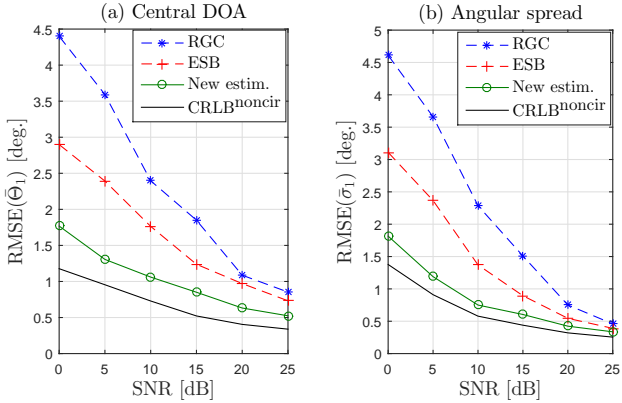}
\caption{RMSE of the three estimators versus $SNR$ for $N=100$, sources with the same angular distributions.}
\label{rmsessnrcrit}
\end{centering}
\end{figure}

In this subsection, the root mean-square error (RMSE) of each estimator is computed empirically by means of $2000$ Monte-Carlo runs. 
We first consider in Fig. \ref{rmsesnap} two uncorrelated ID noncircular sources with the same noncircularity rate ($\gamma_1=\gamma_2=1$) and noncircularity phases $\varphi_1=\frac{\pi}{3}$ and $\varphi_2=\frac{\pi}{4}$. Both ID sources have a Gaussian angular distribution (i.e., GID) and are located at central DOAs $\bar{\Theta}_{1}=10{\textrm{\textdegree}}$ and $\bar{\Theta}_{2}=30{\textrm{\textdegree}}$ with respective angular spreads $\bar{\sigma}_1=1.5{\textrm{\textdegree}}$ and $\bar{\sigma}_2=3{\textrm{\textdegree}}$. The SNR is fixed to $5$ dB while the number of snapshots used to estimate the sample covariance matrix is increased from $100$ to $1000$ in steps of $100$. Figs. \ref{rmsesnap}(a) and \ref{rmsesnap}(b) depict the empirical RMSEs of all tested methods.
Clearly, our estimator is statistically more efficient and outperforms ESB and RGC both in terms of central DOAs and angular spreads estimation accuracy. Moreover, the performance improvements of the proposed method over ESB and RGC hold almost the same irrespectively of $N$. Therefore, we will hereafter fix $N=1000$.\\
Figs. \ref{rmsett1ang}(a) and \ref{rmsett1ang}(b) depict the empirical RMSEs of all tested methods versus the SNR. The analytical RMSE of the new estimator established in Section \ref{section_4} is also plotted there. These figures show a very
good agreement between the empirical and analytical RMSEs of the proposed estimator, thereby corroborating our analytical performance analysis of Section \ref{section_4}. It also suggests that the proposed estimator outperforms ESB and RGC, both in terms of central DOAs and angular spreads estimation capabilities, especially
under the adverse conditions of low SNR levels.\\
In Fig. \ref{rmset1angexp2}, we consider two uncorrelated ID noncircular sources with different angular distributions. More specifically, the first source is uniformly distributed (UID) with central DOA, $\bar{\Theta}_{1}=10{\textrm{\textdegree}}$, and angular spread $\bar{\sigma}_1=1.5{\textrm{\textdegree}}$ while the second is GID distributed with central DOA, $\bar{\Theta}_{2}=30{\textrm{\textdegree}}$, and angular spread $\bar{\sigma}_2=3{\textrm{\textdegree}}$.
To apply ESB in this setup, however, we assume that both sources are GID.
In fact, in contrast to the proposed method and RGC, ESB was specifically derived in the case where all the sources have the same angular distribution.
By comparing Figs. \ref{rmsett1ang} and \ref{rmset1angexp2} (i.e., sources truly having the same distribution),
  we observe that  ESB suffers from severe performance degradation. It even becomes less accurate than RGC at low SNR levels, that is in stark contrast to what was earlier reported in Fig. \ref{rmsett1ang}. The proposed estimator, however, keeps its superiority in terms of estimation accuracy thereby making it more attractive in practice where the sources are more likely to have different angular  distributions.
\\Next, we examine the impact of the sources' separation on the
performance of the three estimators. To that end, we reconsider the case of \textit{noncircular} ID sources with the same angular distribution (GID). The first source is kept fixed at $\bar{\Theta}_1=10{\textrm{\textdegree}}$ with angular spread, $\bar{\sigma}_1=1.5{\textrm{\textdegree}}$, while the second (with  $\bar{\sigma}_2=3^\circ$) is shifted from $18{\textrm{\textdegree}}$ to $30{\textrm{\textdegree}}$ with $2^\circ$. The results are plotted in Fig. \ref{rmset1angexp3} at $5$ dB SNR and suggest that all estimators expectedly improve their accuracy as the DOA separation increases. Yet, the proposed approach  significantly outperforms ESB and RGC  for small DOA separations,  a more challenging scenario in practice.
\\
Finally, we consider in Fig. \ref{rmsessnrcrit} an even more challenging scenario where two uncorrelated GID noncircular sources with the same noncircularity rate ($\gamma_1=\gamma_2=1$) and noncircularity phases $\varphi_1=\frac{\pi}{3}$ and $\varphi_2=\frac{\pi}{4}$ are located at central DOAs $\bar{\Theta}_{1}=10{\textrm{\textdegree}}$ and $\bar{\Theta}_{2}=15{\textrm{\textdegree}}$ with respective angular spreads $\bar{\sigma}_1=2{\textrm{\textdegree}}$ and $\bar{\sigma}_2=4{\textrm{\textdegree}}$. The number of snapshots is fixed to $N=100$. Figs. \ref{rmsessnrcrit}(a) and \ref{rmsessnrcrit}(b) show that the performance of the three methods is satisfactory, especially at high SNR values. However, our new estimator still outperforms the two other methods both in terms of central DOAs and angular spreads estimation performance.
\subsection{Assessment of the new CRLBs: }\label{subsection_6_2}
\begin{figure}[!ht]
\vskip -0.3 cm\hskip -0.2cm
\begin{centering}
\includegraphics[scale=0.53]{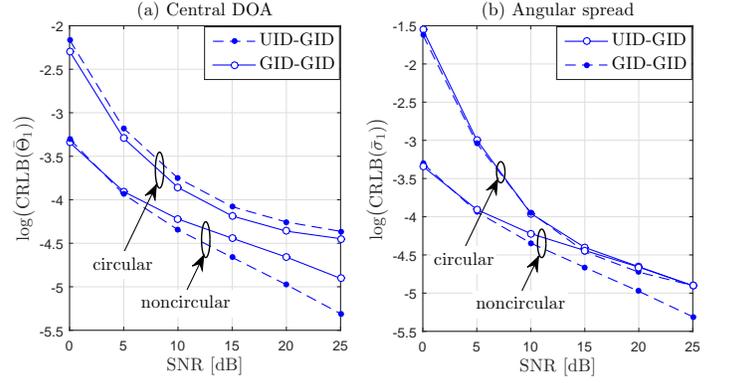}
\vskip -0.1cm
\caption{$\textrm{CRLB}^{\textrm{noncir}}$ and $\textrm{CRLB}^{\textrm{cir}}$
 as function the SNR.}
\label{crb1ang}
\end{centering}
\end{figure}

\begin{figure}[!ht]
\vskip -0.3 cm\hskip -0.2cm
\begin{centering}
\includegraphics[scale=0.53]{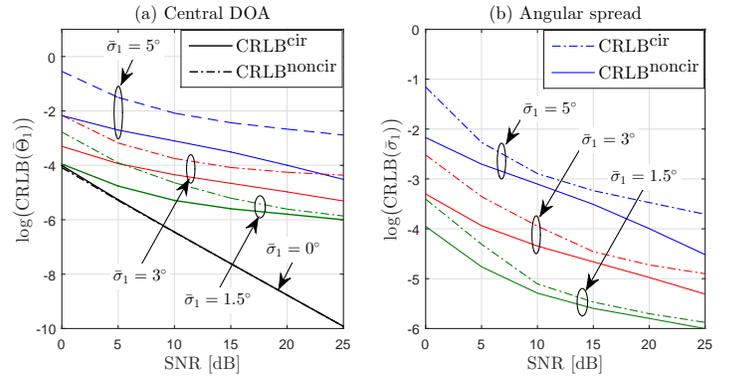}
\vskip -0.1cm
\caption{$\textrm{CRLB}^{\textrm{noncir}}$ and $\textrm{CRLB}^{\textrm{cir}}$  versus the SNR  for different values of $\bar{\sigma}_1$.}
\label{crb1angexp2}
\end{centering}
\end{figure}

\begin{figure}[!ht]
\vskip -0.3 cm\hskip -0.15cm
\begin{centering}
\includegraphics[scale=0.53]{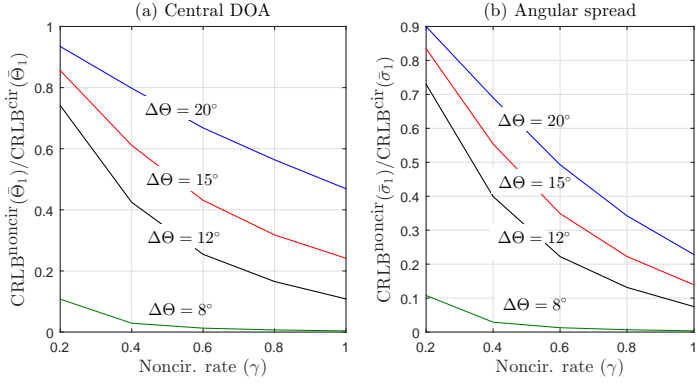}
\vskip -0.1cm
\caption{Ratio of CRLBs as a function of the
noncircularity rate $\gamma$ for different values of DOA separation ($\Delta\Theta$), $\bar{\sigma}_1=3^{\circ}$, $\bar{\sigma}_2=5^{\circ}$, $\varphi_1=\pi/3$, $\varphi_2=\pi/4$, $N=1000$, and SNR $=5$ dB.}
\label{ratioversusrho}
\end{centering}
\end{figure}

\begin{figure}[!ht]
\vskip -0.3 cm\hskip -0.3cm
\begin{centering}
\includegraphics[scale=0.53]{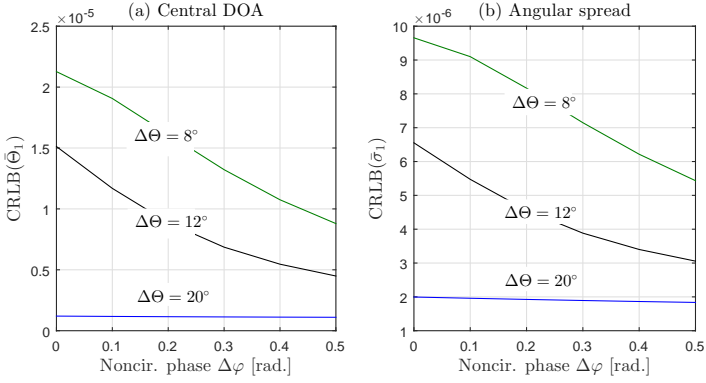}
\vskip -0.1cm
\caption{$\textrm{CRLB}^{\textrm{noncir}}(\bar{\Theta}_1)$ and $\textrm{CRLB}^{\textrm{noncir}}(\sigma_1)$ as a function of the
noncircularity phase $\Delta\varphi$ for different values of DOA separation ($\Delta\Theta$) for $\bar{\sigma}_1=3^{\circ}$, $\bar{\sigma}_2=5^{\circ}$, $\gamma=1$, $N=1000$, and SNR $=5$ dB.}
\label{CRBversusphase}
\end{centering}
\end{figure}

In this subsection, we illustrate the newly derived  CRLBs (i.e., $\textrm{CRLB}^{\textrm{noncir}}$)  in different scenarios. We first consider two equipowered ID sources with identical
noncircularity rate, $\gamma=1$, and noncircularity phases $\varphi_1=\pi/3$ and $\varphi_2=\pi/4$. The sources are located at central DOAs $\bar{\Theta}_{1}=10{\textrm{\textdegree}}$ and $\bar{\Theta}_{2}=30{\textrm{\textdegree}}$ with respective angular spreads $\bar{\sigma}_1=3{\textrm{\textdegree}}$ and $\bar{\sigma}_2=5{\textrm{\textdegree}}$.
Figs. \ref{crb1ang}(a) and \ref{crb1ang}(b) show both $\log(\textrm{CRLB}^{\textrm{noncir}})$ and $\log(\textrm{CRLB}^{\textrm{cir}})$ of $\bar{\Theta}_1$ and $\bar{\sigma}_1$, respectively, when the sources have: $i$)  the same Gaussian angular distribution, and $ii$) different angular distributions (the first source is UID and the second source is GID).
\\
We see from Fig. \ref{crb1ang} that the CRLBs for \textit{noncircular} ID sources are lower than their counterparts derived assuming ID \textit{circular} sources, especially at low SNR values. This illustrates the performance gain that is achieved by exploiting the non-circularity feature of the sources in the estimation process. Moreover, $\textrm{CRLB}^{\textrm{cir}}$ converges faster to $\textrm{CRLB}^{\textrm{noncir}}$, at high SNR, when the sources have the same angular distribution (GID-GID in our case). Therefore, at  high SNRs, the noncircularity of the signals is more informative about the angular parameters when the sources have different distributions.
Next, we examine the impact of the angular spread on the estimation of the angular parameters, by fixing $\bar{\sigma}_2$ and varying $\bar{\sigma}_1$. Fig. \ref{crb1angexp2} depicts $\log(\textrm{CRLB}^{\textrm{noncir}})$ and $\log(\textrm{CRLB}^{\textrm{cir}})$
as a function of the SNR for three different values of $\bar{\sigma}_1$.
Moreover, we consider in Fig. \ref{crb1angexp2}(a) the case of point (or non-distributed) sources which corresponds to $\bar{\sigma}_1=\bar{\sigma}_2=0{\textrm{\textdegree}}$.
As intuitively expected, $\textrm{CRLB}^{\textrm{noncir}}$ and $\textrm{CRLB}^{\textrm{cir}}$ increase with the angular spread and so does the difference between them. This reveals that as the angular spread increases, there is more room for the noncircularity of the signals to
improve the estimation performance. In fact, the signals become more dispersed and thus the \textit{unconjugated} covariance matrix becomes more informative about  the angular parameters.
In Figs. \ref{ratioversusrho} and \ref{CRBversusphase}, we study the effect of the signals' noncircularity parameters on CRLB$^{\textrm{noncir}}$ under different sources' separations,  $\Delta\Theta$, in terms of central DOAs. The first source  is UID and fixed at $\bar{\Theta}_1=10{\textrm{\textdegree}}$  whereas the second source is GID and its  central DOA, $\bar{\Theta}_2$, is varied from $18{\textrm{\textdegree}}$ to $30{\textrm{\textdegree}}$.

We observe from Figs. \ref{ratioversusrho}-(a) and \ref{ratioversusrho}-(b) that $\textrm{CRLB}^{\textrm{noncir}}$ of the two angular parameters decrease as the noncircularity steps rate increases. Moreover,  the gap between $\textrm{CRLB}^{\textrm{noncir}}$ and $\textrm{CRLB}^{\textrm{cir}}$ increases as the DOA separation $\Delta\Theta$ decreases. In fact, the ratio between the two CRLBs tends to zero at low DOA separations (for $\Delta\Theta=8^{\circ}$). More specifically, at low DOA separations, $\textrm{CRLB}^{\textrm{noncir}}$ becomes very small compared to $\textrm{CRLB}^{\textrm{cir}}$, meaning that huge performance gains can be achieved in this challenging scenario by exploiting the additional information carried by the \textit{unconjugated} covariance matrix. Fig. \ref{CRBversusphase} also reveals that $\textrm{CRLB}^{\textrm{noncir}}$ is more sensitive to the noncircularity phase separation at small DOA separations.

\section{Conclusion}  \label{section_7}
In this paper, we developed a new method for the estimation of the angular parameters in the presence of \textit{noncircular} ID sources.
The new estimator decouples the estimation of the central DOAs from that of
the angular spreads by means of two consecutive 1-D searches, thereby resulting in tremendous computational savings as compared to the brute-force 2D grid search solution. It is also oblivious to the sources' angular distribution or any mismatch thereof. This estimator is particulary interessant for symmetric sources' distributions with small angular spreads.\\
The proposed estimator outperforms most recent state-of-the-art techniques, especially for small DOA separations and/or low SNR levels. Its performance was also assessed analytically and the obtained results were corroborated by Monte-Carlo simulations.
In order to benchmark the new estimator, we also derived for the first time an explicit expression for the stochastic CRLBs of the underlying estimation problem. The analysis of the new CRLB unambiguously shows that the noncircularity of the signals brings valuable additional information about the angular parameters especially when the sources have different angular distributions and/or when the angular spreads increase. Besides, the \textit{noncircular} CRLBs decrease as the noncircularity rate increases. And, they are much smaller than the \textit{circular} CRLBs at small DOA separations. In which case they also become more sensitive to the noncircularity phase separation.
\vskip -0.2 cm
\section* {Appendix A---Proof of (\ref{plelemfi})}
From (\ref{conjcovmx}), the $(p,l)$th element of $\mathbf{R}_{ss}(\bar{\bm{\psi}}_k)$ has the following expression:
\begin{eqnarray}
\big[\mathbf{R}_{ss}\big]_{pl}(\bar{\bm{\psi}}_k)&=&\int\rho_k(\theta,\bar{\bm{\psi}}_k)a_p(\theta)a_{l}^{*}(\theta)d\theta,
\end{eqnarray}
where $a_p(\theta)=e^{j2\pi f_{p-1}(\theta)}$.
Otherwise, we denote by $\widetilde{\theta}$ the deviation of the direction $\theta$ from the central DOA $\bar{\Theta}_k$ as follows:
\begin{eqnarray}
\widetilde{\theta}&=&\theta-\bar{\Theta}_k.
\end{eqnarray}
For small angular spreads, $\widetilde{\theta}$ tends to zero. We can therefore use the following approximation:
\begin{eqnarray}\label{apprft}
f_{p-1}(\theta)&\simeq&f_{p-1}(\bar{\Theta}_k)+\widetilde{\theta}f'_{p-1}(\bar{\Theta}_k),
\end{eqnarray}
where $f'_{p-1}(\theta)$ stands for the first derivative of $f_{p-1}(\theta)$ with respect to $\theta$.
Hence, we obtain the following expression for $\big[\mathbf{R}_{ss}\big]_{pl}(\bar{\bm{\psi}}_k)$:
\begin{eqnarray}\label{equcovma}
\!\!\!\!\!\!\!\!\!\!\!\!\big[\mathbf{R}_{ss}\big]_{pl}(\bar{\bm{\psi}}_k)&\!\!\!\!\!\simeq\!\!\!\!\!&e^{j2\pi (f_{p-1}\left(\bar{\Theta}_k)-f_{l-1}(\bar{\Theta}_k)\right)}\nonumber\\
&\!\!\!\!\!\times\!\!\!\!\!&\!\!\int\!\!\!\rho_k(\bar{\Theta}_k+\widetilde{\theta},\bar{\bm{\psi}}_k) e^{j2\pi \left(f'_{p-1}(\bar{\Theta}_k)-f'_{l-1}(\bar{\Theta}_k)\right)\widetilde{\theta}}d\widetilde{\theta}.
\end{eqnarray}
$\big[\mathbf{R}_{ss}\big]_{pl}(\bar{\bm{\psi}}_k)$ can be written equivalently as follows:
\begin{eqnarray}
\big[\mathbf{R}_{ss}\big]_{pl}(\bar{\bm{\psi}}_k)&\!\!\!\simeq\!\!\!&\Big(\mathbf{a}(\bar{\Theta}_k)\!~\mathbf{a}(\bar{\Theta}_k)^H\Big)_{pl}\times \big[\mathbf{T}\big]_{pl}(\bar{\bm{\psi}}_k),
\end{eqnarray}
where $\big[\mathbf{T}\big]_{pl}(\bar{\bm{\psi}}_k)$ is given by:
\begin{eqnarray}\label{plelem}
\!\!\!\!\!\!\!\![\mathbf{T}]_{pl}(\bar{\bm{\psi}}_k)&\!\!\!\!\!=\!\!\!\!\!\!\!&\int \!\!\rho_k(\bar{\Theta}_k+\widetilde{\theta},\bar{\bm{\psi}}_k) e^{j2\pi \left(f'_{p-1}(\bar{\Theta}_k)-f'_{l-1}(\bar{\Theta}_k)\right)\widetilde{\theta}}d\widetilde{\theta}.
\end{eqnarray}
From (\ref{plelem}), the complex conjugate of $[\mathbf{T}]_{pl}(\bar{\bm{\psi}}_k)$ is given by:
\begin{eqnarray}
[\mathbf{T}]^{*}_{pl}(\bar{\bm{\psi}}_k)&\!\!\!=\!\!\!&\int \rho_k(\bar{\Theta}_k+\widetilde{\theta},\bar{\bm{\psi}}_k) e^{-j2\pi \left(f'_{p-1}(\bar{\Theta}_k)-f'_{l-1}(\bar{\Theta}_k)\right)\widetilde{\theta}}d\widetilde{\theta},\nonumber\\
&\!\!\!=\!\!\!&\int \rho_k(\bar{\Theta}_k+\widetilde{\theta},\bar{\bm{\psi}}_k) e^{j2\pi \left(f'_{p-1}(\bar{\Theta}_k)-f'_{l-1}(\bar{\Theta}_k)\right)\widetilde{\theta}}d\widetilde{\theta},\nonumber\\
\label{propreal1}&\!\!\!=\!\!\!&[\mathbf{T}]_{lp}(\bar{\bm{\psi}}_k).
\end{eqnarray}
Moreover, by assuming that the angular distribution is symmetric with respect to the central DOA $\bar{\Theta}_k$, we have the following relation:
\begin{eqnarray}\label{rel_sym}
\rho_k(\theta,\bar{\bm{\psi}}_k)&=&\rho_k(\bar{\Theta}_k+\widetilde{\theta},\bar{\bm{\psi}}_k) =\rho_k(\bar{\Theta}_k-\widetilde{\theta},\bar{\bm{\psi}}_k).
\end{eqnarray}
From (\ref{rel_sym}), it follows that:
\begin{eqnarray}
[\mathbf{T}]^{*}_{pl}(\bar{\bm{\psi}}_k)&\!\!\!\!\!=\!\!\!\!\!\!\!&\int_{-\infty}^{+\infty}\!\!\!\!\rho_k(\bar{\Theta}_k+\widetilde{\theta},\bar{\bm{\psi}}_k) e^{j2\pi \left(f'_{p-1}(\bar{\Theta}_k)-f'_{l-1}(\bar{\Theta}_k)\right)(-\widetilde{\theta})}d\widetilde{\theta},\nonumber\\
&\!\!\!\!\!=\!\!\!\!\!\!\!&\int_{+\infty}^{-\infty} \!\!\!\!\!\!\!\!\!\!\rho_k(\bar{\Theta}_k-\widetilde{\theta_1},\bar{\bm{\psi}}_k) e^{j2\pi \left(f'_{p-1}(\bar{\Theta}_k)-f'_{l-1}(\bar{\Theta}_k)\right)\widetilde{\theta_1}}(-d\widetilde{\theta_1}),\nonumber\\
&\!\!\!\!\!=\!\!\!\!\!\!\!&\int_{-\infty}^{+\infty} \!\!\!\!\rho_k(\bar{\Theta}_k+\widetilde{\theta_1},\bar{\bm{\psi}}_k) e^{j2\pi \left(f'_{p-1}(\bar{\Theta}_k)-f'_{l-1}(\bar{\Theta}_k)\right)\widetilde{\theta_1}}d\widetilde{\theta_1},\nonumber\\
\label{propreal2}&\!\!\!\!\!=\!\!\!\!\!&[\mathbf{T}]_{pl}(\bar{\bm{\psi}}_k).
\end{eqnarray}
From (\ref{propreal1}) and (\ref{propreal2}), we have that:
\begin{eqnarray}
[\mathbf{T}]^{*}_{pl}(\bar{\bm{\psi}}_k)&=&[\mathbf{T}]_{pl}(\bar{\bm{\psi}}_k)=[\mathbf{T}]_{lp}(\bar{\bm{\psi}}_k).
\end{eqnarray}
This proves that $\mathbf{T}(\bar{\bm{\psi}}_k)$ is a real-valued $(L\times L)$  symmetric matrix.
Otherwise, $\exp\{j2\pi \left(f'_{p-1}(\bar{\Theta}_k)-f'_{l-1}(\bar{\Theta}_k)\right)\widetilde{\theta}\}$ can be written as:
\begin{eqnarray}\label{exp_expon}
e^{j2\pi \left(f'_{p-1}(\bar{\Theta}_k)-f'_{l-1}(\bar{\Theta}_k)\right)\widetilde{\theta}}&\!\!\!\!=\!\!\!\!&\cos\!\left(\!2\pi \big(f'_{p-1}(\bar{\Theta}_k)-f'_{l-1}(\bar{\Theta}_k)\big)\widetilde{\theta}\right)\nonumber\\
&\!\!\!\!+\!\!\!\!&j\sin\!\left(\!2\pi \big(f'_{p-1}(\bar{\Theta}_k)-f'_{l-1}(\bar{\Theta}_k)\big)\widetilde{\theta}\right).\nonumber
\end{eqnarray}
Therefore, $[\mathbf{T}]_{pl}(\bar{\bm{\psi}}_k)$ can be expressed as follows:
\begin{eqnarray}
[\mathbf{T}]_{pl}(\bar{\bm{\psi}}_k)&\!\!\!\!\!\!=\!\!\!\!\!\!\!\!&\int\!\! \rho_k(\bar{\Theta}_k+\widetilde{\theta},\bar{\bm{\psi}}_k)\cos\!\left(\!2\pi \big(f'_{p-1}(\bar{\Theta}_k)-f'_{l-1}(\bar{\Theta}_k)\big)\widetilde{\theta}\right)\!d\widetilde{\theta}\nonumber\\
&\!\!\!\!\!\!+\!\!\!\!\!\!&j\!\!\int\!\!\rho_k(\bar{\Theta}_k+\widetilde{\theta},\bar{\bm{\psi}}_k)\sin\!\left(\!2\pi \big(f'_{p-1}(\bar{\Theta}_k)-f'_{l-1}(\bar{\Theta}_k)\big)\widetilde{\theta}\right)\!d\widetilde{\theta}.\nonumber
\end{eqnarray}
Since $\mathbf{T}(\bar{\bm{\psi}}_k)$ is a real-valued matrix, then we can deduce that:
\begin{eqnarray}
\!\!\!\!\!\!\!\!\!\!\int\!\! \rho_k(\bar{\Theta}_k+\widetilde{\theta},\bar{\bm{\psi}}_k)\sin\!\left(\!2\pi \big(f'_{p-1}(\bar{\Theta}_k)-f'_{l-1}(\bar{\Theta}_k)\big)\widetilde{\theta}\right)\!d\widetilde{\theta}&\!\!\!\!\!\!=\!\!\!\!\!\!&0.
\end{eqnarray} 
Consequently, $[\mathbf{T}]_{pl}(\bar{\bm{\psi}}_k)$ can be reduced to: 
\begin{eqnarray}
[\mathbf{T}]_{pl}(\bar{\bm{\psi}}_k)&\!\!\!\!\!\!=\!\!\!\!\!\!\!\!&\int \rho_k(\bar{\Theta}_k+\widetilde{\theta},\bar{\bm{\psi}}_k)\!\cos\!\left(2\pi \big(f'_{p-1}(\bar{\Theta}_k)\!-\!f'_{l-1}(\bar{\Theta}_k)\big)\widetilde{\theta}\right)\!d\widetilde{\theta},\nonumber\\
&\!\!\!\!\!\!=\!\!\!\!\!\!\!\!&\int\!\!\rho_k(\theta,\bar{\bm{\psi}}_k)\!\cos\!\left(2\pi (f'_{p-1}(\bar{\Theta}_k)\!-\!f'_{l-1}(\bar{\Theta}_k))(\theta\!-\!\bar{\Theta}_k)\right)\!d\theta,\nonumber
\end{eqnarray}
thereby leading to the result given in (\ref{plelemfi}).
Moreover, since, we have:
\begin{eqnarray}
\cos\!\left(2\pi (f'_{p-1}(\bar{\Theta}_k)\!-\!f'_{l-1}(\bar{\Theta}_k))(\theta\!-\!\bar{\Theta}_k)\right)&\leq&1,
\end{eqnarray}
then we can conclude that $\forall p, l$, we have:
\begin{eqnarray}
[\mathbf{T}]_{pl}(\bar{\bm{\psi}}_k)&\leq&\int \rho_k(\theta,\bar{\bm{\psi}}_k)d\theta=1.
\end{eqnarray}
\vskip -0.2 cm
\section* {Appendix B---Proof of (\ref{cost_function1})}
Substituting (\ref{newexpextfm}) in (\ref{newcrit}) and using the identity $\textrm{tr}\{\mathbf{A}\mathbf{B}\mathbf{C}\}=\textrm{tr}\{\mathbf{C}\mathbf{A}\mathbf{B}\}$ for any square matrices, $\mathbf{A}, \mathbf{B}$ and $\mathbf{C}$, along with the fact that $\widetilde{\bm{\Phi}}(\bar{\Theta},\varphi)^H\widetilde{\bm{\Phi}}(\bar{\Theta},\varphi)=\mathbf{I}_{2L}$, we obtain:
\begin{eqnarray}\label{newexpminfonc}
\!\!\!\!\!\!\!\!\!\!f\big(\bm{\psi},\varphi\!~\big|\!~\widehat{\mathbf{R}}_{\widetilde{\mathbf{x}}\widetilde{\mathbf{x}}}^{-2}\big)&=&\textrm{tr}\left\{\widetilde{\bm{\Phi}}(\bar{\Theta},\varphi)\widetilde{\mathbf{T}}^2(\bm{\psi})\widetilde{\bm{\Phi}}(\bar{\Theta},\varphi)^H\widehat{\mathbf{R}}_{\widetilde{\mathbf{x}}\widetilde{\mathbf{x}}}^{-2}\right\}.
\end{eqnarray}
On the other hand, by recalling the expression of the extended array response vector, $\widetilde{\mathbf{a}}(\bar{\Theta},\varphi)$, in (\ref{extended-response}), it  follows that $\widetilde{\bm{\Phi}}(\bar{\Theta},\varphi)=\textrm{diag}\{\widetilde{\mathbf{a}}(\bar{\Theta},\varphi)\}$ is given by:
\begin{eqnarray}\label{expdith}
\!\!\!\!\!\!\!\!\!\widetilde{\bm{\Phi}}(\bar{\Theta},\varphi)&\!\!=\!\!&\left(\!\!\begin{array}{ccc}
\textrm{diag}\big\{\mathbf{a}(\bar{\Theta})\big\}&~~& \bm{0}_{L\times L}\\\\
\bm{0}_{L\times L}&~~& e^{-j\varphi}\textrm{diag}\big\{\mathbf{a}(\bar{\Theta})\big\}^H
\end{array}\!\!\right)\!\!.
\end{eqnarray}
Furthermore, by recalling the expression of $\widetilde{\mathbf{T}}(\bm{\psi})$  in (\ref{ext_T}), it is easy to show that:
\begin{eqnarray}\label{expfinalT2}
\widetilde{\mathbf{T}}(\bm{\psi})^2&=&\left(\!\!\begin{array}{cc}
\mathbf{A}(\bm{\psi})~~~~~~\mathbf{B}(\bm{\psi})\\
\mathbf{B}(\bm{\psi})~~~~~\mathbf{A}(\bm{\psi})
\end{array}\!\!\right),
\end{eqnarray}
where
\begin{eqnarray}
\mathbf{A}(\bm{\psi})&=&\mathbf{T}(\bm{\psi})^2~+~{\mathbf{T}'}(\bm{\psi})^2,\nonumber\\
\mathbf{B}(\bm{\psi})&=&\mathbf{T}(\bm{\psi}){\mathbf{T}'}(\bm{\psi})~+~{\mathbf{T}'}(\bm{\psi})\mathbf{T}(\bm{\psi}).\nonumber
\end{eqnarray}
Similar to (\ref{eigextR}), the estimated extended covariance matrix, $\widehat{\mathbf{R}}_{\widetilde{\mathbf{x}}\widetilde{\mathbf{x}}}$, is eigendecomposed as follows:
\begin{eqnarray}
\!\!\!\!\!\!\!\!\widehat{\mathbf{R}}_{\widetilde{\mathbf{x}}\widetilde{\mathbf{x}}}&=&\widehat{\widetilde{\mathbf{U}}}_s\widehat{\bm{\Sigma}}_s\widehat{\widetilde{\mathbf{U}}}_s^H~+~\widehat{\widetilde{\mathbf{U}}}_w\widehat{\bm{\Sigma}}_w\widehat{\widetilde{\mathbf{U}}}_w^H,
\end{eqnarray}
from which it can be shown that:
\begin{eqnarray}
\label{expRxinvv}\widehat{\mathbf{R}}_{\widetilde{\mathbf{x}}\widetilde{\mathbf{x}}}^{-2}&=&\widehat{\widetilde{\mathbf{U}}}_s\widehat{\bm{\Sigma}}_s^{-2}\widehat{\widetilde{\mathbf{U}}}_s^H~+~\widehat{\widetilde{\mathbf{U}}}_w\widehat{\bm{\Sigma}}_w^{-2}\widehat{\widetilde{\mathbf{U}}}_w^H.
\end{eqnarray}
Moreover, as shown in [\ref{LEEE32}], $\widehat{\widetilde{\mathbf{U}}}_s$ and $\widehat{\widetilde{\mathbf{U}}}_w$ can be partitioned as follows:
\begin{eqnarray}
\label{rel1}\!\!\!\!\!\!\!\!\mathsmaller{\widehat{\widetilde{\mathbf{U}}}_s}&=&[\mathsmaller{\widehat{\mathbf{U}}_s^T},\mathsmaller{\widehat{\mathbf{U}}_s^{'T}}]^T~~~~~~~ \textrm{with}~~~~~~~ \widehat{\mathbf{U}}_s^{'}~=~\widehat{\mathbf{U}}_{s}^*\mathbf{D}_s,\\
\label{rel2}\!\!\!\!\!\!\!\!\mathsmaller{\widehat{\widetilde{\mathbf{U}}}_w}&=&[\mathsmaller{\widehat{\mathbf{U}}_{w}^T},\mathsmaller{\widehat{\mathbf{U}}_w^{'T}}]^T~~~~~~~ \textrm{with}~~~~~~~ \widehat{\mathbf{U}}_w^{'}~=~\widehat{\mathbf{U}}_{w}^*\mathbf{D}_w,
\end{eqnarray}
%
%
where $\mathbf{D}_s$ and $\mathbf{D}_w$ are some diagonal matrices whose complex diagonal entries are of unit modulus.
Injecting (\ref{rel1}) and (\ref{rel2}) back into (\ref{expRxinvv}), we show after some algebraic manipulations that $\widehat{\mathbf{R}}_{\widetilde{\mathbf{x}}\widetilde{\mathbf{x}}}^{-2}$ has the following block diagonal structure:
\begin{eqnarray}\label{newexpRinvv}
\widehat{\mathbf{R}}_{\widetilde{\mathbf{x}}\widetilde{\mathbf{x}}}^{-2}&=&\left(\!\!
\begin{array}{cc}
\widehat{\mathbf{R}}_1~~~~~\widehat{\mathbf{R}}_2\\
\widehat{\mathbf{R}}_2^*~~~~~\widehat{\mathbf{R}}_1^*
\end{array}\!\!\right),
\end{eqnarray}
where
\begin{eqnarray}
\widehat{\mathbf{R}}_1&=&\widehat{\mathbf{U}}_{s}\widehat{\bm{\Sigma}}^{-2}_s\widehat{\mathbf{U}}_{s}^H~~+~~\widehat{\mathbf{U}}_{w}\widehat{\bm{\Sigma}}_w\widehat{\mathbf{U}}_{w}^H,\\
\widehat{\mathbf{R}}_2&=&\widehat{\mathbf{U}}_{s}\widehat{\bm{\Sigma}}^{-2}_s\mathbf{D}_s^*\widehat{\mathbf{U}}_{s}^T~~+~~\widehat{\mathbf{U}}_{n}\widehat{\bm{\Sigma}}_w\mathbf{D}_w^*\widehat{\mathbf{U}}_{w}^T.
\end{eqnarray}
Substituting (\ref{expdith}), (\ref{expfinalT2}), and (\ref{newexpRinvv}) back into (\ref{newexpminfonc}), and resorting to  some algebraic manipulations yields the following result:
\begin{eqnarray}\label{cost_function1}
\!\!\!\!\!\!\!\!\!\!f\big(\bm{\psi},\varphi\big|\!~\widehat{\mathbf{R}}_{\widetilde{\mathbf{x}}\widetilde{\mathbf{x}}}^{-2}\big)&=&2\Re\Big\{~\!z_1(\bm{\psi})~\!+~\!e^{j\varphi}z_2(\bm{\psi})~\!\Big\}~,
\end{eqnarray}
in which the complex numbers $z_1(\bm{\psi})$ and $z_2(\bm{\psi})$ are explicitly given by:
\begin{eqnarray}
\!\!\!\!\!\!\!\!\label{z_1}z_1(\bm{\psi})&=&\textrm{tr}\Big\{\textrm{diag}\{\mathbf{a}(\Theta)\}\mathbf{A}(\bm{\psi})\textrm{diag}\left\{\mathbf{a}(\Theta)^H\right\}\widehat{\mathbf{R}}_1\Big\}, \\
\!\!\!\!\!\!\!\!\label{z_2}z_2(\bm{\psi})&=&\textrm{tr}\Big\{\textrm{diag}\{\mathbf{a}(\Theta)\}\mathbf{B}(\bm{\psi})\textrm{diag}\left\{\mathbf{a}(\Theta)^T\right\}\widehat{\mathbf{R}}_2^*\Big\}.
\end{eqnarray}
In order to reduce the dimensionality of the optimization problem at hand, we begin by minimizing the underlying cost function with respect to the unknown noncircularity phase $\varphi$. To that end, we use $z_2(\bm{\psi})=|z_2(\bm{\psi})|\mathlarger{\exp\{j\angle z_2(\bm{\psi})}\}$ and  rewrite (\ref{cost_function1}) as follows:
\begin{eqnarray}
\!\!\!\!\!\!\!f\big(\bm{\psi},\varphi\!~\big|\!~\widehat{\mathbf{R}}_{\widetilde{\mathbf{x}}\widetilde{\mathbf{x}}}^{-2}\big)
&&\nonumber\\
&&\!\!\!\!\!\!\!\!\!\!\!\!\!\!\!\!\!\!\!\!\!\!\!\!\!\!\!\!\!=~2\Re\big\{z_1(\bm{\psi})\big\}~\!+~\!2|z_2(\bm{\psi})|\Re\big\{\mathlarger{e^{j\angle z_2(\bm{\psi})}}\mathlarger{e^{j\varphi}}~\!\Big\}~,\\
\label{cost_function21}&&\!\!\!\!\!\!\!\!\!\!\!\!\!\!\!\!\!\!\!\!\!\!\!\!\!\!\!\!\!=~2\Re\big\{z_1(\bm{\psi})\big\}~\!+~\!2|z_2(\bm{\psi})|\cos\big(\varphi+\angle z_2(\bm{\psi})\big).
\end{eqnarray}
From (\ref{cost_function21}), it is clear  (for a fixed $\bm{\psi}$) that the function $f(.)$ attains its minimum (with respect to $\varphi$) at the point:
\begin{eqnarray}\label{phi_est}
\widehat{\varphi}(\bm{\psi})&=&\pi~-~\angle~\!z_2(\bm{\psi}).
\end{eqnarray}
Substituting (\ref{phi_est}) back into (\ref{cost_function21}) and recalling (\ref{z_1}) and (\ref{z_2}), we obtain the following cost function that depends on $\bm{\psi}$ only:
\begin{eqnarray}
\!\!\!\!\!\!\!\!\!\!\!\!f_c\big(\bm{\psi}\!~\big|\!~\widehat{\mathbf{R}}_{\widetilde{\mathbf{x}}\widetilde{\mathbf{x}}}^{-2}\big)&\!\!\!\!&\nonumber\\&&\!\!\!\!\!\!\!\!\!\!\!\!\!\!\!\!\!\!\!\!\!\!\!\!=~\!\Re\bigg\{\textrm{tr}\Big\{\textrm{diag}\big\{\mathbf{a}(\Theta)\big\}\mathbf{A}(\bm{\psi})\textrm{diag}\big\{\mathbf{a}(\Theta)^H\big\}\widehat{\mathbf{R}}_1\Big\}\bigg\}\nonumber\\
&\!\!\!\!&\!\!\!\!\!\!-\bigg|~\!\textrm{tr}\Big\{\textrm{diag}\big\{\mathbf{a}(\Theta)\big\}\mathbf{B}(\bm{\psi})\textrm{diag}\big\{\mathbf{a}(\Theta)\big\}^T\widehat{\mathbf{R}}_2^*\Big\}\bigg|,
\end{eqnarray}
which is equivalent to the result given in (\ref{expgpsi}).
\section* {Appendix C---Proof of (\ref{constraint_1}) and (\ref{constraint_2_compact})}
$\mathbf{T}(\bm{\psi})$ is a symmetric Toeplitz matrix constructed from its first column vector $\mathbf{t}_1$ as follows:
\begin{eqnarray}
\mathbf{T}(\bm{\psi})&=&\textrm{Toeplitz}\big(\mathbf{t}_1\big),
\end{eqnarray}
where the $l$th element of the vector $\mathbf{t}_1$ is given from (\ref{simpexpT}) by:
\begin{eqnarray}
\!\!\!\!\mathbf{t}_1(l)&\!\!\!\!\!\!=\!\!\!\!\!\!&[\mathbf{T}]_{l1}(\bar{\bm{\psi}}_k),\nonumber\\
&\!\!\!\!\!\!=\!\!\!\!\!\!&\int\! \rho_k(\theta,\bar{\bm{\psi}}_k)\cos\left(2\pi(l-1)g(\bar{\Theta}_k)\!(\theta\!-\!\bar{\Theta}_k)\right)\!d\theta.
\end{eqnarray}
For small angular spreads, we use a second-order Taylor-series development of $\cos\left(2\pi(l-1)g(\bar{\Theta}_k)\!(\theta\!-\!\bar{\Theta}_k)\right)$ to obtain the following equality:
\begin{eqnarray}
\cos\!\left(2\pi(l-1)g(\bar{\Theta}_k))\!(\theta\!-\!\bar{\Theta}_k)\right)&\!\!\!\!\!\simeq\!\!\!\!\!&1-2\pi^2(l-1)^2g^2(\bar{\Theta}_k))\!(\theta\!-\!\bar{\Theta}_k)^2.\nonumber
\end{eqnarray}
Then, $\mathbf{t}_1(l)$ can be approximated as follows:
\begin{eqnarray}
\mathbf{t}_1(l)&\!\!\!\!\!\simeq\!\!\!\!\!\!\!&\int\!\! \rho_k(\theta,\bar{\bm{\psi}}_k)d\theta-2\pi^2(l-1)^2g^2(\bar{\Theta}_k)\!\!\int\! (\theta\!-\!\bar{\Theta}_k)^2\rho_k(\theta,\bar{\bm{\psi}}_k)d\theta,\nonumber\\
\label{newexpfirvec}&\!\!\!\!\!=\!\!\!\!\!&1-2\pi^2(l-1)^2g^2(\bar{\Theta}_k)\bar{\sigma}_{k}^2.
\end{eqnarray}
From (\ref{newexpfirvec}), we clearly see that  the elements, $\{\mathbf{t}_1(l)\}_{l=1}^{L}$, of the vector,  $\mathbf{t}_1$, satisfy the following property:
\begin{equation}
1~=~\mathbf{t}_1(1)~\geq~ \mathbf{t}_1(2)~\geq\ldots \geq~\mathbf{t}_1(L).
\end{equation}
Moreover, if $\sigma<1/\big(\sqrt{2}\pi(l-1)g(\bar{\Theta}_k)\big)~\forall ~\!l=1,\ldots,L$ or equivalently $\sigma<1/\big(\sqrt{2}\pi(L-1)g(\bar{\Theta}_k)\big)$, then $\mathbf{t}_1(L)\geq 0$.\\
In the same way, ${\mathbf{T}'}(\bm{\psi})$ is a Hankel matrix defined from its first and last column vectors $\mathbf{t}'_1$ and $\mathbf{t}'_L$ as follows:
\begin{eqnarray}
\mathbf{T}'(\bm{\psi})&=&\textrm{Hankel}\big(\mathbf{t}'_1,\mathbf{t}'_L\big),
\end{eqnarray}
where the $l$th elements of the vectors $\mathbf{t}'_1$ and $\mathbf{t}'_L$ are given, respectively, by:
\begin{eqnarray}
\!\!\!\!\!\!\!\!\mathbf{t}'_1(l)&\!\!\!\!\!=\!\!\!\!\!&[\mathbf{T}']_{l1}(\bar{\bm{\psi}}_k),\nonumber\\
&\!\!\!\!\!=\!\!\!\!\!&\int\!\! \rho_k(\theta,\bar{\bm{\psi}}_k)\cos\!\left(2\pi(l+1-2)g(\bar{\Theta}_k)\!(\theta\!-\!\bar{\Theta}_k)\right)\!d\theta,\nonumber\\
&\!\!\!\!\!=\!\!\!\!\!&\int\!\! \rho_k(\theta,\bar{\bm{\psi}}_k)\cos\!\left(2\pi(l-1)g(\bar{\Theta}_k)\!(\theta\!-\!\bar{\Theta}_k)\right)\!d\theta,\\
\!\!\!\!\!\!\!\!\mathbf{t}'_L(l)&\!\!\!\!\!=\!\!\!\!\!&[\mathbf{T}']_{lL}(\bar{\bm{\psi}}_k),\nonumber\\
&\!\!\!\!\!=\!\!\!\!\!&\int\!\! \rho_k(\theta,\bar{\bm{\psi}}_k)\cos\!\left(2\pi(L+l-2)g(\bar{\Theta}_k)\!(\theta\!-\!\bar{\Theta}_k)\right)\!d\theta.
\end{eqnarray}
For small angular spreads, $\mathbf{t}'_1(l)$ and $\mathbf{t}'_L(l)$ can be approximated as follows:
\begin{eqnarray}
\label{newexpfirvecpr1}\mathbf{t}'_1(l)&\!\!\!\!\!\simeq\!\!\!\!\!&1-2\pi^2(l-1)^2g^2(\bar{\Theta}_k)\bar{\sigma}_{k}^2,\\
\label{newexpfirvecpr2}\mathbf{t}'_L(l)&\!\!\!\!\!\simeq\!\!\!\!\!&1-2\pi^2(L+l-2)^2g^2(\bar{\Theta}_k)\bar{\sigma}_{k}^2.
\end{eqnarray}
From (\ref{newexpfirvecpr1}) and (\ref{newexpfirvecpr2}), we see clearly that the elements of $\mathbf{t}'_1$ and $\mathbf{t}'_L$ satisfy the following properties:
\begin{eqnarray}
1~=~\mathbf{t}'_1(1)~&\geq&~ \mathbf{t}'_1(2)~\geq\ldots \geq~\mathbf{t}'_1(L),\\
\mathbf{t}'_L(1)~&\geq&~ \mathbf{t}'_L(2)~\geq\ldots \geq~\mathbf{t}'_L(L).\\
\mathbf{t}'_1(L)~&=&~\mathbf{t}'_L(1).
\end{eqnarray}
Moreover, if $\sigma<1/\big(2\sqrt{2}\pi(L-1)g(\bar{\Theta}_k)\big)$, then $\mathbf{t'}_L(L)\geq 0$.
\section* {Appendix D---Proof of (\ref{additional_bias_00})}
 For any small vector and matrix perturbations, $\delta\mathbf{x}$ and  $\delta\mathbf{X}$,  and scalar-valued function $g(\mathbf{x},\mathbf{X})$, we have the following Taylor series expansion [\ref{LEEE36}, \ref{LEEE37}] around $\mathbf{x}_0$ and  $\mathbf{X}_0$:
\begin{eqnarray}\label{result_0AppB}
g(\mathbf{x}_0+\delta\mathbf{x},\mathbf{X}_0+\delta\mathbf{X})&&\nonumber\\
&&\!\!\!\!\!\!\!\!\!\!\!\!\!\!\!\!\!\!\!\!\!\!\!\!\!\!\!\!\!\!\!\!\!\!\!\!\!\!\!\!\!\!\!\!\!\!\!\!\!\!\!\!\!\!=~\!g(\mathbf{x}_0,\mathbf{X}_0)+  \frac{\partial g}{\partial \mathbf{x}}(\mathbf{x}_0,\mathbf{X}_0)^T\delta\mathbf{x}+\textrm{tr}\left\{\frac{\partial g}{\partial \mathbf{X}}(\mathbf{x}_0,\mathbf{X}_0)^T\delta\mathbf{X}\right\}.\nonumber\\
&&
\end{eqnarray}
The result in (\ref{result_0AppB}) is applied  with  $\mathbf{x}=\bm{\alpha}$ and  $\mathbf{X}=\mathbf{R}$ to the functions:
\begin{eqnarray}\label{f_alpha}
f_{\alpha_i}(\bm{\alpha}|\mathbf{R})&\triangleq&\partial f(\bm{\alpha}|\mathbf{R})/\partial\alpha_i,~~~~~~i=1,2,3
\end{eqnarray}
 in order to obtain their Taylor series expansions around the point $(\mathbf{x}_0,\mathbf{X}_0)=(\breve{\bm{\alpha}}_k,\breve{\mathbf{R}}_{\widetilde{\mathbf{x}}\widetilde{\mathbf{x}}}^{-2})$. In this way, for  $k=1,2,\ldots K$, the underlying perturbations are given by:
 \begin{eqnarray}
 \delta\mathbf{x}=\bm{\alpha}-\breve{\bm{\alpha}}_k~~~~~~\textrm{and}~~~~~~\delta\mathbf{X}=\mathbf{R}-\breve{\mathbf{R}}_{\widetilde{\mathbf{x}}\widetilde{\mathbf{x}}}^{-2}.
 \end{eqnarray}
 By doing so, we obtain for $i=1,2,3$:
 \begin{eqnarray}\label{element_wise_appB}
 f_{\alpha_i}(\bm{\alpha}|\mathbf{R})&\!\!=\!\!&  f_{\alpha_i}\big(\breve{\bm{\alpha}}_k|\breve{\mathbf{R}}_{\widetilde{\mathbf{x}}\widetilde{\mathbf{x}}}^{-2}\big)~+ \left[\frac{\partial f_{\alpha_i}}{\partial\bm\alpha}(\breve{\bm{\alpha}}_k|\breve{\mathbf{R}}_{\widetilde{\mathbf{x}}\widetilde{\mathbf{x}}}^{-2})\right]^T\!\!(\bm{\alpha}-\breve{\bm{\alpha}}_k)\nonumber\\
&&~~~~~~~~~~~~~~~~~~~~~~~~~~~~~~~~~~~+~v_i(\breve{\bm{\alpha}}_k~\!|~\!\breve{\mathbf{R}}_{\widetilde{\mathbf{x}}\widetilde{\mathbf{x}}}^{-2},\mathbf{R}),\nonumber\\
 \end{eqnarray}
 where
 \begin{eqnarray}\label{v_i}
\!\!\!\!\!\!\!\!\!\!\!\!v_i(\breve{\bm{\alpha}}_k~\!|~\!\breve{\mathbf{R}}_{\widetilde{\mathbf{x}}\widetilde{\mathbf{x}}}^{-2},\mathbf{R})&\!\!\!=\!\!\!&\textrm{tr}\left\{\left[\frac{\partial f_{\alpha_i}}{\partial\mathbf{R}}\!\!\left(\breve{\bm{\alpha}}_k\big|\breve{\mathbf{R}}_{\widetilde{\mathbf{x}}\widetilde{\mathbf{x}}}^{-2}\right)\right]^T\!\!\!\big(\mathbf{R}-\breve{\mathbf{R}}_{\widetilde{\mathbf{x}}\widetilde{\mathbf{x}}}^{-2}\big)\right\}\!.
\end{eqnarray}
Moreover, notice from (\ref{gradient_definition}) and (\ref{Hessian_definition}) that $f_{\alpha_i}(\bm{\alpha}|\mathbf{R})$ and $\left[\partial f_{\alpha_i}(\bm{\alpha}|\mathbf{R})/\partial\bm\alpha\right]^T$ are, respectively, the $i^{th}$ element of the gradient vector,  $\mathbf{f}(\bm{\alpha}|\mathbf{R})$, and   the $i^{th}$ row of the Hessian matrix $\mathbf{F}(\bm{\alpha}|\mathbf{R})$.  Therefore, by further defining the vector $\mathbf{v}=[v_1,v_2,v_3]^T$, the results of (\ref{element_wise_appB}) for $i=1,2,3$ are rewritten in the following matrix/vector form:
\begin{eqnarray}\label{result1_appB}
\mathbf{f}(\bm{\alpha}|\mathbf{R})&\!\!\!\!\!\!&\nonumber\\
&&\!\!\!\!\!\!\!\!\!\!\!\!\!\!\!\!\!\!\!\!\!\!=~\!\mathbf{f}(\breve{\bm{\alpha}}_k|\breve{\mathbf{R}}_{\widetilde{\mathbf{x}}\widetilde{\mathbf{x}}}^{-2})~+~ \mathbf{F}(\breve{\bm{\alpha}}_k|\breve{\mathbf{R}}_{\widetilde{\mathbf{x}}\widetilde{\mathbf{x}}}^{-2})(\bm{\alpha}-\breve{\bm{\alpha}}_k)~+~\mathbf{v}(\breve{\bm{\alpha}}_k|\breve{\mathbf{R}}_{\widetilde{\mathbf{x}}\widetilde{\mathbf{x}}}^{-2},\mathbf{R}).\\
\end{eqnarray}
Evaluating the  expansion in (\ref{result1_appB}) at $(\bm{\alpha},\mathbf{R})=\big(\widehat{\bm{\alpha}}_k$, $\widehat{\mathbf{R}}_{\widetilde{\mathbf{x}}\widetilde{\mathbf{x}}}^{-2}\big)$ and using $\Delta\breve{\bm{\alpha}}_k\triangleq\widehat{\bm{\alpha}}_k-\breve{\bm{\alpha}}_k$ leads to:
\begin{eqnarray}\label{result1_appB1}
\mathbf{f}(\widehat{\bm{\alpha}}_k|\widehat{\mathbf{R}}_{\widetilde{\mathbf{x}}\widetilde{\mathbf{x}}}^{-2})&\!\!\!\!\!\!&\nonumber\\
&&\!\!\!\!\!\!\!\!\!\!\!\!\!\!\!\!\!\!\!\!\!\!\!\!\!\!\!\!=~\!\mathbf{f}(\breve{\bm{\alpha}}_k|\breve{\mathbf{R}}_{\widetilde{\mathbf{x}}\widetilde{\mathbf{x}}}^{-2})~+~ \mathbf{F}(\breve{\bm{\alpha}}_k|\breve{\mathbf{R}}_{\widetilde{\mathbf{x}}\widetilde{\mathbf{x}}}^{-2})\Delta\breve{\bm{\alpha}}_k~+~\mathbf{v}\big(\breve{\bm{\alpha}}_k|\breve{\mathbf{R}}_{\widetilde{\mathbf{x}}\widetilde{\mathbf{x}}}^{-2},\widehat{\mathbf{R}}_{\widetilde{\mathbf{x}}\widetilde{\mathbf{x}}}^{-2}\big).\nonumber\\
\end{eqnarray}
The finite-sample and asymptotic estimates, $\widehat{\bm{\alpha}}_k$ and $\breve{\bm{\alpha}}_k$, are obtained by minimizing $f(\bm{\alpha}|\widehat{\mathbf{R}}_{\widetilde{\mathbf{x}}\widetilde{\mathbf{x}}}^{-2})$ and $f(\bm{\alpha}|\breve{\mathbf{R}}_{\widetilde{\mathbf{x}}\widetilde{\mathbf{x}}}^{-2})$, respectively. Therefore,  the gradient of  the latter objective function is identically zero at $\widehat{\bm{\alpha}}_k$ and $\breve{\bm{\alpha}}_k$, i.e.:
\!\!\!\!\!\!\begin{eqnarray}\label{gradient_null}
f(\widehat{\bm{\alpha}}_k|\widehat{\mathbf{R}}_{\widetilde{\mathbf{x}}\widetilde{\mathbf{x}}}^{-2})~=~\mathbf{0}_3~~~~~~~\textrm{and}~~~~~~~f(\breve{\bm{\alpha}}_k|\breve{\mathbf{R}}_{\widetilde{\mathbf{x}}\widetilde{\mathbf{x}}}^{-2})~=~\mathbf{0}_3.
\end{eqnarray}
Exploiting (\ref{gradient_null}) back into (\ref{result1_appB1}) and resolving for $\Delta\breve{\bm{\alpha}}_k$, one obtains:
\begin{eqnarray}\label{additional_bias_0}
\Delta\breve{\bm{\alpha}}_k&=&\mathbf{F}^{-1}(\breve{\bm{\alpha}}_k|\mathbf{R}_{\widetilde{\mathbf{x}}\widetilde{\mathbf{x}}}^{-2})~\!\mathbf{v}(\breve{\bm{\alpha}}_k|\mathbf{R}_{\widetilde{\mathbf{x}}\widetilde{\mathbf{x}}}^{-2},\widehat{\mathbf{R}}_{\widetilde{\mathbf{x}}\widetilde{\mathbf{x}}}^{-2}),
\end{eqnarray}
 in which owing to (\ref{relainvinvR}) we also replaced $\breve{\mathbf{R}}_{\widetilde{\mathbf{x}}\widetilde{\mathbf{x}}}^{-2}$ by $\mathbf{R}_{\widetilde{\mathbf{x}}\widetilde{\mathbf{x}}}^{-2}$.
To find the explicit expression of $\mathbf{v}(\breve{\bm{\alpha}}_k|\mathbf{R}_{\widetilde{\mathbf{x}}\widetilde{\mathbf{x}}}^{-2},\widehat{\mathbf{R}}_{\widetilde{\mathbf{x}}\widetilde{\mathbf{x}}}^{-2})$, involved  we  further denote:
\begin{eqnarray}\label{DeltaR1}
\Delta\mathbf{R}_{\widetilde{\mathbf{x}}\widetilde{\mathbf{x}}}^{-2}&\triangleq&\widehat{\mathbf{R}}_{\widetilde{\mathbf{x}}\widetilde{\mathbf{x}}}^{-2}~-~\mathbf{R}_{\widetilde{\mathbf{x}}\widetilde{\mathbf{x}}}^{-2},
\end{eqnarray}
Then, using  (\ref{gradient_definition}) and (\ref{f_alpha}) in  (\ref{v_i}), it follows that:
\begin{eqnarray}
\!\!\!\!\!\!\!\!\!\!\!\!\!\!\!v_i(\breve{\bm{\alpha}}_k|\mathbf{R}_{\widetilde{\mathbf{x}}\widetilde{\mathbf{x}}}^{-2},\widehat{\mathbf{R}}_{\widetilde{\mathbf{x}}\widetilde{\mathbf{x}}}^{-2})&\!\!\!\!\!\!&\nonumber \\
&&\!\!\!\!\!\!\!\!\!\!\!\!\!\!\!\!\!\!\!\!\!\!\!\!=\!~\textrm{tr}\left\{\left[\frac{\partial}{\partial\mathbf{R}}\textrm{tr}\left\{\mathbf{R}~\!\widetilde{\mathbf{R}}^{[i]}_{ss}\right\}\right]^T\!\!\!\Delta\mathbf{R}_{\widetilde{\mathbf{x}}\widetilde{\mathbf{x}}}^{-2}\right\}\Bigg|_{\substack{\!\!\bm{\alpha}\!~=\!~\breve{\bm{\alpha}}_k\\~\mathbf{R}\!~=\!~\mathbf{R}_{\widetilde{\mathbf{x}}\widetilde{\mathbf{x}}}^{-2}}}\!\!\!,
\end{eqnarray}

where $\widetilde{\mathbf{R}}^{[i]}_{ss}$ is given by (\ref{R_i}).

\section* {Appendix E---Derivation of $\textrm{CRLB}(\bm{\eta})$}
We have the following parameter vector:
\begin{eqnarray}
\bm{\upsilon}&=&\left[\bm{\eta}^T,~\bm{\xi}^T\right]^T\!.\nonumber
\end{eqnarray}
Therefore, the associated FIM can be written as:
\begin{eqnarray}\label{blockmatr}
\mathbf{I}(\bm{\upsilon})&=&\left(
\begin{array}{cc}
\mathbf{I}_{\bm{\eta},\bm{\eta}}&\mathbf{I}_{\bm{\xi},\bm{\eta}}\\
\mathbf{I}_{\bm{\eta},\bm{\xi}}&\mathbf{I}_{\bm{\xi},\bm{\xi}}
\end{array}\right),
\end{eqnarray}
whose $ij$th entry is expressed as:
\begin{eqnarray}
\left[\mathbf{I}(\bm{\upsilon})\right]_{ij}&=&\mathsmaller{\frac{N}{2}}\textrm{tr}\left\{\frac{\partial\mathbf{R}_{\widetilde{\mathbf{x}}\widetilde{\mathbf{x}}}}{\partial\upsilon_i}\mathbf{R}_{\widetilde{\mathbf{x}}\widetilde{\mathbf{x}}}^{-1}\frac{\partial\mathbf{R}_{\widetilde{\mathbf{x}}\widetilde{\mathbf{x}}}}{\partial\upsilon_j}\mathbf{R}_{\widetilde{\mathbf{x}}\widetilde{\mathbf{x}}}^{-1}\right\},
\end{eqnarray}
with
\begin{eqnarray}\label{eqpartR}
\frac{\partial\mathbf{R}_{\widetilde{\mathbf{x}}\widetilde{\mathbf{x}}}}{\partial\upsilon_i}&=&\left(
\begin{array}{cc}
\frac{\partial\mathbf{R}_{\mathbf{x}\mathbf{x}}}{\partial\upsilon_i}&\frac{\partial\mathbf{R}'_{\mathbf{x}\mathbf{x}}}{\partial\upsilon_i}\\\\
\left(\frac{\partial\mathbf{R}'_{\mathbf{x}\mathbf{x}}}{\partial\upsilon_i}\right)^*&\left(\frac{\partial\mathbf{R}_{\mathbf{x}\mathbf{x}}}{\partial\upsilon_i}\right)^*
\end{array}\right).
\end{eqnarray}
In (\ref{eqpartR}), $\upsilon_i$ is the $i$th element of $\bm{\upsilon}$ and the involved partial derivatives of $\mathbf{R}_{\mathbf{x}\mathbf{x}}$ are given by:
\begin{eqnarray}
\frac{\partial\mathbf{R}_{\mathbf{x}\mathbf{x}}}{\partial\bar{\Theta}_i}&=&\sigma_{s_i}^2\left(\frac{\partial\bm{\Phi}}{\partial\bar{\Theta}_i}\mathbf{T}\bm{\Phi}^H+\bm{\Phi}\frac{\partial\mathbf{T}}{\partial\bar{\Theta}_i}\bm{\Phi}^H+\bm{\Phi}\mathbf{T}\frac{\partial\bm{\Phi}^H}{\partial\bar{\Theta}_i}\right),\nonumber\\
\frac{\partial\mathbf{R}_{\mathbf{x}\mathbf{x}}}{\partial\sigma_i}&=&\sigma_{s_i}^2\bm{\Phi}\frac{\partial\mathbf{T}}{\partial\sigma_i}\bm{\Phi}^H,\nonumber\\
\frac{\partial\mathbf{R}_{\mathbf{x}\mathbf{x}}}{\partial\sigma_{s_i}^2}&=&\bm{\Phi}\mathbf{T}{\Phi}^H,\nonumber\\
\frac{\partial\mathbf{R}_{\mathbf{x}\mathbf{x}}}{\partial\sigma_n^2}&=&\mathbf{I}_L,\nonumber\\
\frac{\partial\mathbf{R}_{\mathbf{x}\mathbf{x}}}{\partial\varphi_i}&=&\bm{0}_{L\times L}.\nonumber
\end{eqnarray}
Furthermore, it can be shown that the partial derivatives of $\mathbf{R}'_{\mathbf{x}\mathbf{x}}$ are given by:
\begin{eqnarray}
\frac{\partial\mathbf{R}'_{\mathbf{x}\mathbf{x}}}{\partial\bar{\Theta}_i}&=&\sigma_{s_i}^2 e^{j\varphi_i}\left(\frac{\partial\bm{\Phi}}{\partial\bar{\Theta}_i}\mathbf{T}'\bm{\Phi}^T+\bm{\Phi}\frac{\partial\mathbf{T}'}{\partial\bar{\Theta}_i}\bm{\Phi}^T+\bm{\Phi}\mathbf{T}'\frac{\partial\bm{\Phi}^T}{\partial\bar{\Theta}_i}\right),\nonumber\\
\frac{\partial\mathbf{R}'_{\mathbf{x}\mathbf{x}}}{\partial\sigma_i}&\!\!\!\!\!=\!\!\!\!\!&\sigma_{s_i}^2 e^{j\varphi_i}\bm{\Phi}\frac{\partial\mathbf{T}'}{\partial\sigma_i}\bm{\Phi}^T,\nonumber\\
\frac{\partial\mathbf{R}'_{\mathbf{x}\mathbf{x}}}{\partial\sigma_{s_i}^2}&\!\!\!\!\!=\!\!\!\!\!&e^{j\varphi_i}\bm{\Phi}\mathbf{T}'{\Phi}^T,\nonumber\\
\frac{\partial\mathbf{R}'_{\mathbf{x}\mathbf{x}}}{\partial\sigma_n^2}&\!\!\!\!\!=\!\!\!\!\!&\bm{0}_{L\times L},\nonumber\\
\frac{\partial\mathbf{R}'_{\mathbf{x}\mathbf{x}}}{\partial\varphi_i}&\!\!\!\!\!=\!\!\!\!\!&j\sigma_{s_i}^2 e^{j\varphi_i}\bm{\Phi}\mathbf{T}'{\Phi}^T.\nonumber
\end{eqnarray}

Recall that our goal is to find the CRLB of the angular parameters, $\bm{\eta}$, denoted as $\textrm{CRLB}(\bm{\eta})$. Therefore, we are interested in the $\bm{\eta}$-block of
$\mathbf{I}^{-1}(\bm{\upsilon})$ only.
From (\ref{blockmatr}), the whole FIM, $\mathbf{I}(\bm{\upsilon})$, is a block matrix with $\mathbf{I}_{\bm{\eta},\bm{\eta}}$ being its first diagonal block.
Thus, we use the block matrices  inversion
Lemma [\ref{LEEE39}] to obtain the following expression for $\textrm{CRLB}(\bm{\eta})$:
\begin{eqnarray}
\textrm{CRLB}(\bm{\eta})&=&\left(\mathbf{I}_{\bm{\eta},\bm{\eta}}~-~\mathbf{I}_{\bm{\xi},\bm{\eta}}^T\mathbf{I}_{\bm{\xi},\bm{\xi}}^{-1}\mathbf{I}_{\bm{\xi},\bm{\eta}}\right)^{-1}.
\end{eqnarray}

\end{document}